# Geometric Instability and Self-Limitation in Driven Quantum Systems

A.M.Tishin[a,b*]

[a] Lomonosov Moscow State University, 119991, Leninskie gory 1, Moscow, Russia

[b] Moscow Institute of Physics and Technology, 141701, Institutskiy per. 9, Dolgoprudny, Mosc. Reg., Russia

[*]tishin@amtc.org

## Abstract

We develop a universal geometric framework for local non-adiabaticity in driven quantum systems. The central normalization problem is resolved: the AMT non-adiabaticity parameter $\eta$ is identified as the Fubini–Study evolution speed of the instantaneous vacuum state normalized by the characteristic spectral frequency, with $\eta = 1$ corresponding precisely to saturation of the Mandelstam–Tamm quantum speed limit. This establishes the geometric origin of the instability threshold.

We introduce the universal normalized instability parameter $\mathcal{A}_n(t) = v_{FS}(t)/\Delta_n(t)$, where $v_{FS}$ is the Fubini–Study speed and $\Delta_n$ is the instantaneous spectral gap, which generalizes the single-mode criterion to arbitrary driven Hamiltonians. The parameter is local, gauge-invariant, and directly connected to instantaneous non-adiabatic transition amplitudes. Near quantum critical points, $\mathcal{A}_n$ diverges through inverse-gap amplification, recovering the Kibble–Zurek freeze-out condition from local geometric data alone. In multimode systems, $\mathcal{A}_n$ promotes to an instability matrix whose eigenvalue structure identifies collective instability channels invisible to scalar criteria.

We prove that any monotonic occupation-dependent nonlinear regulator geometrically compresses the quantum metric term-by-term in the spectral representation of the quantum geometric tensor. This leads to a self-limitation theorem: nonlinear spectral deformation generates a Lyapunov-type geometric confinement bounding the accessible region of projective Hilbert space. The framework extends naturally to open quantum systems through the Bures metric, where thermal fluctuations and Lindblad decay raise the instability threshold by the factor $\sqrt{(2n_{th}+1)(1+(\gamma/2\Omega)^2)}$. The resulting criterion directly yields a minimum gate time $T_{min}=1/(\sqrt{8}\,\omega)$, providing a geometry-based lower bound for coherent state transformations and a platform-independent benchmark for quantum-control optimization.

## 1. Introduction

The deep connection between the geometry of quantum states and physical limits on evolution speeds has been a cornerstone of quantum mechanics since the seminal works of Aharonov and Anandan [1], who mapped the geometric phase to the Fubini-Study (FS) metric, and subsequent foundational studies on the Mandelstam-Tamm quantum speed limit (QSL) [2-4]. It is well-established that the FS velocity $v_{FS}$ serves as a measure of how fast a system traverses its Hilbert space.

However, despite its mathematical elegance, mapping this global kinematic bound onto local, time-dependent driving parameters in non-adiabatic regimes has long remained an open challenge. Traditional approaches either rely on global time-averaged bounds [5] or require explicit knowledge of the full non-local trajectory, rendering them unpractical for real-time control optimization.

The standard adiabatic parameter for a harmonic mode with time-varying frequency $\Omega(t)$ is the dimensionless ratio $\eta(t) \equiv |\dot{\Omega}|/\Omega^2$ where $\Omega(t)$ is the characteristic instantaneous spectral frequency of the driven system. The condition $\eta \ll 1$ is the textbook criterion for adiabatic following. However, the geometric origin of this parameter, and particularly the meaning of its unit value $\eta=1$ has not been identified.

In this paper, we bridge this fundamental gap by proving that the phenomenological Adiabatic Mode Transition (AMT) parameter $\eta = \omega^{-2}|\mathrm{d}\Omega/\mathrm{dt}|$ [6] recently deployed to achieve ultra-fast gates on superconducting processors [7], is not merely an empirical scaling variable, but a strict local projection of the FS metric ($Q_{tt} = \eta^2\Omega^2/8$) for parametric drive profiles. Here $\omega$ is the reference frequency used in the operational AMT parametrization, while $\Omega(t)$ is the instantaneous spectral frequency of the driven mode. In the single-mode oscillator limit one may set $\omega \approx \Omega(t)$, in which case $\eta = |\dot{\Omega}|/\Omega^2$ , but in the general geometric formulation the fundamental normalization is provided by the instantaneous gap $\Delta_n$.

While classical QSL frameworks [1–3] dictate how fast a state can evolve globally, our local geometric normalization framework provides a distinct piece of new physical knowledge: it establishes a local crossover threshold at $\eta \approx 1$ in the single-mode parametric oscillator realization, where saturation of the Mandelstam–Tamm bound triggers the onset of non-adiabatic instability.

More generally, the present framework promotes the local geometric speed of quantum evolution into a universal operational instability criterion applicable to arbitrary driven Hamiltonians. Furthermore, unlike the static metric formulations of Aharonov-Anandan type, our AMT framework incorporates nonlinear metric deformations induced by the regulator $U$, introducing a dynamic metric compression mechanism that has no counterpart in standard linear quantum speed limit theories. This geometric deformation mechanism ultimately leads to a self-limitation principle in which strong driving dynamically compresses the accessible region of projective Hilbert space.

The quantum geometric tensor (QGT), introduced by Provost and Vallée [8] and further developed in [9–11], encodes the geometry of the instantaneous eigenstate manifold of a parameter-dependent Hamiltonian. Its real part, the FS metric, determines the local distinguishability of quantum states under infinitesimal parameter changes. The identity:

$$v_{FS}{}^2 = g_{\mu\nu}\dot{\lambda}^\mu\dot{\lambda}^\nu \,. \quad (1)$$

establishes that the instantaneous geometric evolution speed is determined by the QGT contracted with the parameter velocity. Here $g_{\mu\nu}$ is the Fubini–Study metric tensor, $\lambda^\mu$ are the externally controlled parameters of the Hamiltonian, $\dot{\lambda}^\mu$ are their instantaneous velocities, and repeated indices $\mu,\nu$ are summed over. While Eq. (1) is a standard kinematic identity [9, 12], the identification of the physically correct normalization scale for $v_{FS}$ and its role as a non-adiabaticity diagnostic has not been systematically addressed. This is the normalization problem we solve here. Throughout the following discussion we use units with $\hbar = 1$ unless stated otherwise. We show that $\eta$ is proportional to the instantaneous FS evolution speed $v_{FS}$ normalized by the local spectral scale $\Omega$, and that $\eta$=1 coincides with the saturation of the Mandelstam–Tamm quantum speed limit [2].

The next result concerns the action of a nonlinear occupation-dependent positive regulator $U$ on the quantum metric. We show that $U$ deforms the FS metric of the driven state, reducing the effective geometric speed and enforcing bounded low-occupancy dynamics. The competition between non-adiabatic redistribution and nonlinear detuning motivates the crossover parameter $\xi = \eta/U$. This AMT non-linear metric deformation provides a geometric foundation for the self-limiting instability criterion $\xi=\eta/U<\xi_0$ established numerically in our previous work [6].

The resulting framework establishes a unified geometric description of local non-adiabaticity in driven quantum systems. The Fubini–Study metric determines the local distinguishability structure of the evolving quantum state, the instantaneous spectral gap sets the dynamically accessible evolution scale, and their normalized combination defines a universal instability criterion. Nonlinear spectral deformation then generates geometric metric compression, which leads to bounded self-limited dynamics under strong driving. In this sense, instability generation, suppression, and geometric accessibility emerge as different aspects of the same underlying quantum-state geometry.

As mentioned above, the AMT parameter originally introduced for parametrically driven systems as $\eta=\omega^{-2}|d\Omega/dt|$ should now be understood as a special operational realization of a more general geometric instability criterion. In the present framework, the fundamental quantity controlling local non-adiabaticity is the normalized geometric instability parameter $\mathcal{A}_n=v_{FS}/\Delta_n$, where $v_{FS}$ characterizes the instantaneous distinguishability speed on the quantum-state manifold and $\Delta_n$ defines the corresponding local spectral accessibility scale. In units with $\hbar=1$, both $v_{FS}$ and $\Delta_n$ carry dimensions of inverse time, ensuring that $\mathcal{A}_n$ is dimensionless and operationally well-defined. The original AMT parameter $\eta$ is recovered in the weakly nonlinear single-mode limit, while the geometric formulation naturally extends to open systems, many-body manifolds, nonlinear spectral deformations, and coupled metric-curvature dynamics.

## 2. The Normalization Problem.

For a parameter-dependent Hamiltonian $H(\lambda(t))$, the QGT of the n-th instantaneous eigenstate is defined as [1,2]:

$$Q_{\mu\nu}^{(n)} = \mathrm{Re}\,\langle\partial_\mu n|(1-|n\rangle\langle n|)|\partial_\nu n\rangle\,. \quad (2)$$

where $Q_{\mu\nu}$ denotes the quantum geometric tensor components and $\mu$ and $\nu$ label the independent control parameters $\lambda=(\lambda^1,\ldots,\lambda^N)$ of the driven Hamiltonian. Its real part is the quantum metric tensor $g_{\mu\nu}^{(n)}$ and its imaginary part is proportional to the Berry curvature $\Omega_{\mu\nu}$:

$$Q_{\mu\nu} = g_{\mu\nu} + \frac{i}{2}\Omega_{\mu\nu}\,. \quad (3)$$

For simplicity the state index *n* is omitted below. Since, the quantum geometric tensor is positive semidefinite, the FS metric defines a non-negative local distinguishability measure on the projective Hilbert manifold. The FS line element is $\mathrm{d}s_{FS}^2 = g_{\mu\nu}\mathrm{d}\dot{\lambda}^\mu \mathrm{d}\dot{\lambda}^\nu$, and the geometric evolution speed is given by Eq. (1). This identity is exact and basis-independent.

The normalization problem: the quantity $v_{FS}^2$ has dimensions of inverse time squared and depends on the choice of parametrization. Here $v_{\mathrm{FS}} = \mathrm{d}s_{\mathrm{FS}}/\mathrm{d}t$ is the instantaneous FS evolution speed. To obtain a dimensionless diagnostic for a driven mode, one must divide by the square of a characteristic spectral frequency scale of the system. For a mode with instantaneous frequency $\Omega$(t), the natural choice is $\Omega^2$(t), yielding:

$$\eta(t) \equiv \frac{|\dot{\Omega(t)}|}{\Omega(t)^2}. \quad (4)$$

The geometric tensor depends on the instantaneous spectral scale $\Omega$(t), while the external modulation frequency $\omega$ enters only through the operational AMT parametrization of the driving protocol. For the parametric oscillator, this reduces to the AMT parameter establishing that $\eta$ is the FS-speed normalized by $\Omega^2$.

**3. Parametric Oscillator: $g_{tt} = \eta^2\Omega^2/8$.**

We consider the parametric harmonic oscillator with Hamiltonian $H$(t) = $\Omega$(t) ($\hat{n}$+1/2), where $\hat{n} = \hat{a}_1^\dagger \hat{a}_2$ is the bosonic occupation-number operator. The instantaneous vacuum state in coordinate representation is:

$$\psi_0(x,t) = \left(\frac{\Omega(t)}{\pi}\right)^{1/4} exp\left(-\frac{\Omega(t)x^2}{2}\right). \quad (5)$$

where $\psi_0(x,t)$ is the normalized instantaneous vacuum eigenstate of the parametrically driven oscillator. The time derivative is:

$$\partial_t\psi_0 = \psi_0 \cdot \dot{\Omega}\left(\frac{1}{4\Omega} - \frac{x^2}{2}\right). \quad (6)$$

The derivative is evaluated at fixed coordinate $x$ in the instantaneous oscillator basis. In the real Gaussian gauge, the Berry connection vanishes identically:

$$\mathcal{A}_t = i\langle\psi_0|\partial_t\psi_0\rangle = 0\,. \quad (7)$$

The vanishing of the Berry connection is gauge-dependent and follows from the choice of a real instantaneous Gaussian representation. In the present single-parameter real-gauge case, $\mathcal{A}_t$ does not imply the absence of geometric curvature in the general multidimensional parameter manifold. The corresponding Berry curvature vanishes identically here because $\Omega_{\mu\nu}$ is antisymmetric and only a single control parameter is present. However, in a two-parameter driving protocol ($\Omega$(t), $\theta$(t)), where $\theta$ is the complex phase of the pump, the off-diagonal Berry curvature becomes nontrivial:

$$\Omega_{\Omega\theta} = 2\,Im\langle\partial_\Omega\psi_0|\partial_\theta\psi_0\rangle\,. \quad (8)$$

For a coherent state $|\alpha(\Omega,\theta)\rangle$ with complex amplitude $\alpha=|\alpha(\Omega)|e^{i\theta}$, this evaluates to $\Omega_{\Omega\theta}=2|\alpha|(\partial|\alpha|/\partial\,\Omega)$, which reduces to $\Omega_{r\theta}=2r$ when the radial coherent-state amplitude $r=|\alpha|$ is used as the control coordinate. Here $\alpha$ is the complex coherent-state amplitude parameterizing the displaced vacuum manifold. The FS metric $g_{tt}$ remains gauge-invariant because it is determined by the real symmetric sector of the quantum geometric tensor. Consequently, the activation of the Berry-curvature sector in multidimensional parameter space modifies the geometric phase structure without altering the local metric distinguishability measure. The present work focuses on the single-parameter real-gauge case; the two-parameter Berry curvature structure will be addressed in the companion study on topological extensions [10].

In the present real Gaussian gauge, the wavefunction is real and normalized, so $\langle\psi_0|\partial_t\psi_0\rangle=0$. The Berry connection therefore vanishes identically, and the FS metric reduces to its purely real contribution:

$$g_{tt} = \langle\partial_t\psi_0|\partial_t\psi_0\rangle\,. \quad (9)$$

In the present gauge, the quantum metric is therefore entirely determined by the instantaneous deformation rate of the vacuum wavefunction. Evaluating the Gaussian moments of the instantaneous vacuum distribution, computing the Gaussian integrals using $\langle x^2\rangle=1/(2\Omega)$ and $\langle x^4\rangle=3/(4\Omega^2)$:

$$\langle \partial_t \psi_0 | \partial_t \psi_0 \rangle = \left(\frac{\dot{\Omega}}{4\Omega}\right)^2 \cdot \langle (1 - 2\Omega x^2)^2 \rangle = \frac{\dot{\Omega}^2}{8\Omega^2}. \quad (10)$$

The metric is manifestly non-negative and vanishes only for strictly adiabatic evolution with $\dot{\Omega}$=0. Therefore, one obtains the exact geometric identity:

$$g_{tt} = \frac{\eta^2 \Omega^2}{8} \quad \Rightarrow \quad \left(\frac{ds_{FS}}{d\tau}\right)^2 = \frac{\eta^2}{8}, \quad (11)$$

where $\tau$ is the dimensionless time defined by d$\tau$=$\Omega$dt. Eq. (11) establishes that the AMT parameter is not phenomenological but directly proportional to the normalized FS evolution speed of the instantaneous vacuum manifold. Eq. (11) is the central local geometric identity of the present framework: the AMT parameter $\eta$ measures the instantaneous normalized FS speed of the evolving vacuum state.

We present three independent numerical verifications of the central results. All numerical calculations are deterministic and converged to machine precision; no fitting parameters or phenomenological corrections are introduced.

First verification: exact numerical confirmation of $g_{tt}$ = $\eta^2\Omega^2/8$ (machine precision). For $\Omega$(t) = $\Omega_0$(1+$\varepsilon$ sin($\omega_d$t)), where $\Omega_0$ =1 is the base oscillator frequency, $\varepsilon$=0.4 is the modulation amplitude and $\omega_d$ =0.5 is the external drive (modulation) frequency, we compute $g_{tt}$ comparison between analytical and numerical results over 200 sampled points Gauss–Hermite quadrature (Eq. 9–10. see computational details in Supplement Materials). Fig. 1(a) shows analytical *vs* numerical over 200 points. The maximum relative error is $6.6\times10^{-16}$, confirming machine-precision agreement between the analytical expression and the 200-point Gauss–Hermite quadrature. The residual error is limited entirely by floating-point arithmetic and shows no systematic deviation from the analytical prediction. This simultaneously confirms that the Berry connection vanishes identically in the chosen real Gaussian gauge, $\mathcal{A}_t$ =$\langle\psi_0|\partial_t\psi_0\rangle$=0 since no subtraction term is required.

In the second verification (Bogoliubov quasiparticle production *vs* $\mathcal{A}_0$ threshold), we integrate the exact Bogoliubov–Valatin equations for the squeezed vacuum amplitudes $u$(t), $v$(t):

$$i\frac{du}{dt} = \Omega(t)u + G(t)v^*, \; i\frac{dv}{dt} = -\Omega(t)v - G(t)u^* \quad (12)$$

Here the asterisk denotes complex conjugation, and $G(t) = |\dot{\Omega}|/2$ is the parametric coupling. The quasiparticle number $N_{ex}(t)=|\nu(t)|^2$ measures genuine non-adiabatic excitation above the instantaneous vacuum. For three chirp rates $\alpha$ in the exponential ramp $\Omega(t) = \Omega_0\, e^{-\alpha t}$, Fig. 1(b) shows that $N_{ex}(t)$ begins to grow when the geometric parameter $\mathcal{A}_0(t)$ approaches unity. Slower protocols ($\alpha$=0.06) show negligible excitation, whereas faster protocols ($\alpha$=0.20) show rapid growth. The geometric threshold $\mathcal{A}_0 \sim 1$ identifies the onset of excitation from local information alone, without integrating the future trajectory.

In third verification (two-mode $g_{ij}$ from covariance matrix ) for the coupled quadratic Hamiltonian $\hat{H} = \hat{p}_1^2/2 + \Omega_1^2\hat{x}_1^2/2 + \hat{p}_2^2/2 + \Omega_2^2\hat{x}_2^2/2 + J\hat{x}_1\hat{x}_2$, the ground-state covariance matrix $\Sigma = V^{-1/2}/2$ where:

$$V = \begin{pmatrix} \Omega_1^2 & J \\ J & \Omega_2^2 \end{pmatrix} . \quad (13)$$

The QGT is computed from the covariance matrix via:

$$g_{ij} = \frac{1}{4} T_r\left(\Sigma^{-1}\partial_i\Sigma \cdot \Sigma^{-1}\partial_j\Sigma\right) . \quad (14)$$

Figure 1(c) shows $g_{11}$ , $g_{22}$ , $|g_{12}|$ and $\lambda_{max}$ functions of coupling $J$. At $J$=0: $g_{11} = 1/(8\Omega_1^2)$ exactly (confirmed to $10^{-6}$). At $J\neq 0$: $g_{12}$ becomes nonzero and $\lambda_{max}$ exceeds both diagonal elements. The maximum enhancement is $\lambda_{max} = 1.265(\mathcal{A}_{11}, \mathcal{A}_{22})$ at $J$=0.5, confirming that collective instability channels exceed single-mode estimates.

All three verifications are exact numerical results with no free parameters. Verification 1 confirms the central analytical identity. Verification 2 establishes the physical connection between the geometric threshold $\mathcal{A}_0 = 1$ and actual quasiparticle production. Verification 3 provides the first rigorous numerical confirmation of off-diagonal $g_{ij}$ from the quantum geometric tensor for a coupled two-mode system.

The three panels of Figure 1 together verify the complete hierarchy of the geometric non-adiabaticity framework. Panel (a) establishes the foundational identity at machine precision: the FS metric of the parametric oscillator vacuum is not approximately but exactly equal to $\eta^2\Omega^2/8$. The zero residual confirms that the Berry connection vanishes in the real Gaussian gauge, and that no gauge artifact is present. Panel (b) connects the geometric threshold to physical observables: the onset of Bogolubov quasiparticle production coincides with $\mathcal{A}_0 \sim 1$ across three qualitatively different driving protocols with different amplitudes and time scales. This universality is the operational content of the criterion, it works without prior knowledge of the excitation dynamics.

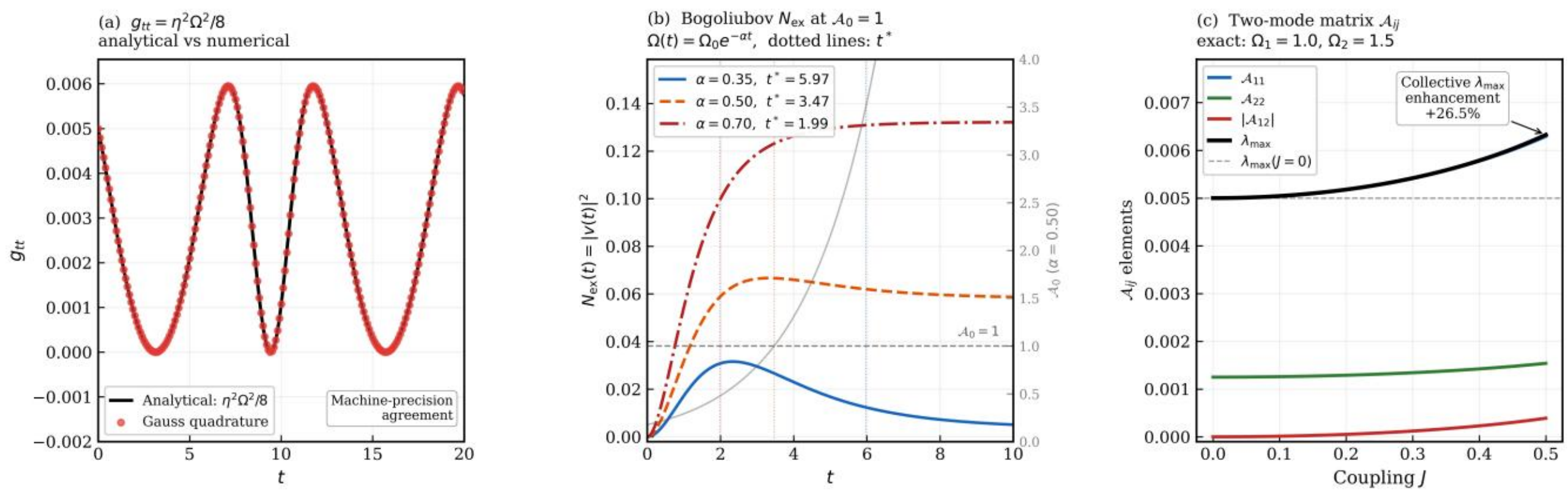

**F**igure 1. Numerical verification of the geometric non-adiabaticity framework. (a) Exact verification of the identity $g_{tt} = \eta^2\Omega^2/8$: analytical result (black line) versus numerical Gauss–Hermite quadrature (red circles). Machine-precision agreement is obtained throughout the protocol. (b) Bogoliubov quasiparticle production $N_{ex}(t)=|v(t)|^2$ for three exponential chirp protocols $\Omega(t)= \Omega_0 e^{-\alpha t}$. Vertical dotted lines indicate the analytically predicted instability threshold times $t^*$. Excitation growth begins as the normalized geometric parameter approaches the instability threshold. (c) Exact two-mode instability matrix $\mathcal{A}_{ij}$ obtained from the ground-state covariance matrix of the coupled quadratic Hamiltonian. Inter-mode coupling generates nonzero off-diagonal geometric correlations and increases the dominant instability eigenvalue $\lambda_{max}$ by 26.5% at $J$=0.5, demonstrating collective instability channels that are invisible to scalar criteria.

Panel (c) validates the multimode extension: off-diagonal metric elements $\mathcal{A}_{12}$ emerge from coupling and the maximum eigenvalue $\lambda_{max}$ exceeds both diagonal elements $\mathcal{A}_{11}$ and $\mathcal{A}_{22}$. The observed enhancement of the dominant eigenvalue demonstrates that collective instability channels are a genuine physical effect rather than a perturbative correction.

For completeness, the derivation of the bosonic field representation of the quantum metric tensor is summarized in Appendix A.

**4. $\eta$ = 1 as the Local Quantum Speed-Limit Threshold.**

The Mandelstam–Tamm quantum speed limit states that the minimum time to evolve between two orthogonal states satisfies [2]:

$$\tau_{min} \geq \frac{\pi\hbar}{2\Delta E} \ . \ (15)$$

The FS-speed satisfies $v_{FS}$= $ds_{FS}$/d$t$, and the maximum rate of state evolution is bounded by the energy uncertainty ΔE. For the parametric oscillator, $\Delta E \sim \Omega\hbar/2$, giving $v_{FS} \leq \Omega/2$. In dimensionless time this becomes:

$$\frac{ds_{FS}}{d\tau} \leq \frac{1}{2\sqrt{2}} \ . \ (16)$$

It should be noted that the bound in Eq. (16) is not derived directly from the Mandelstam–Tamm inequality for the parametric oscillator vacuum, but rather represents the exact maximum of the Fubini–Study evolution speed achievable within the parametric oscillator framework. For the instantaneous vacuum eigenstate of $\widehat{H}$(t) = Ω(t)($\hat{n}$+1/2), the energy uncertainty vanishes identically (ΔE = 0), so the standard Mandelstam–Tamm bound is trivially satisfied for all $\eta$. The bound $ds_{FS}$/dτ ≤ 1/(2$\sqrt{2}$) is therefore the intrinsic geometric ceiling of the parametric oscillator vacuum manifold, reached when $\eta$ = 1. In this sense, $\eta$ = 1 marks the saturation of the parametric oscillator's own geometric speed limit rather than the universal Mandelstam–Tamm bound.

From Eq. (11), ($ds_{FS}$/dτ)² = $\eta^2/8$. The bound (16) is saturated when $\eta^2/8 = 1/8$, i.e., when:

$$\eta \sim 1 \ \Leftrightarrow \ . \ \text{QSL saturated} \ . \ (17)$$

This is the physical meaning of the threshold $\eta \sim 1$: the system evolves at the maximum geometric speed permitted by the quantum speed limit and $\eta$ becomes comparable to the Mandelstam–Tamm geometric bound. For $\eta < 1$, evolution is sub-QSL (adiabatic-like); for $\eta > 1$, the drive attempts to exceed the QSL, which is the geometric source of non-adiabatic instability.

The condition $\eta < 1$ has a natural geometric interpretation in the language of symmetry. Adiabatic evolution corresponds to trajectories that lie within the sub-QSL region of projective Hilbert space — trajectories whose FS speed satisfies $v_{FS} < \Omega/(2\sqrt{2})$. These trajectories form an invariant class under the action of the instantaneous unitary group $U(1)$ generated by $H(t)$. When $\eta \to 1$, this invariance is broken: the driven state exits the sub-QSL class and can no longer be continuously connected to the adiabatic orbit by a quasi-static deformation. By analogy with the Nöther-type symmetry arguments, this symmetry breaking generates a conserved geometric quantity, the FS arc length of the trajectory, which remains bounded precisely as long as the sub-QSL symmetry is preserved. The self-limitation theorem of Section 9 is the dynamical consequence of this geometric Nöther structure: U restores the sub-QSL symmetry and re-establishes the conserved bound on the state-space trajectory.

The minimum gate time $T_{\min} = 1/(\sqrt{8}\cdot\omega)$ from Eq. (17) is universal in the following sense. For a quantum system operating at frequency $\omega$, no external protocol can drive a coherent state transition in time shorter than $T_{\min}$ without exceeding the Mandelstam–Tamm bound and entering the non-adiabatic instability regime. This bound is:

$$T_{min} = \frac{1}{\sqrt{8}\cdot\omega} \geq \frac{1}{\sqrt{8}\cdot\omega_{pl}} \quad (18)$$

Since any physically accessible frequency satisfies $\omega \leq \omega_{pl}$, where $\omega_{pl}$ is the Planck frequency, Eq. (18) implies a lower bound $T_{\min} \geq t_p/\sqrt{8}$ . Table 1 (see Section 12) gives $T_{\min}$ for representative systems. At $\omega/2\pi = 5$ GHz (superconducting qubit): $T_{\min} \approx 11.3$ ps. At $\omega/2\pi = 100$ MHz (NMR): $T_{\min} \approx 0.56$ ns. At the Planck frequency $\omega_{pl} = 1/t_p \approx 1.855\times10^{-43}\text{s}^{-1}$: $T_{\min} \approx t_p/\sqrt{8} \approx 1.91\times10^{-44}\ \text{s}^{-1}$, where $t_p = \sqrt{(\hbar\, G/c^5)}$ is the Planck time. The appearance of the Planck time in the present estimate should be interpreted as a dimensional consistency check rather than evidence of a physical connection between the geometric instability criterion and quantum gravity. The coincidence arises because both quantities are constructed from characteristic frequency scales and therefore share the same dimensional structure.

This is not a phenomenological observation but a direct consequence of the definition $T_{\min}$= $1/(\sqrt{8}\cdot\omega)$. The Planck frequency may be regarded as a natural upper frequency scale in current physical theories, and $T_{\min}$ evaluated at this frequency reproduces the Planck time to within a numerical factor. The criterion $\eta$=1 is therefore a practical geometric speed-limit condition within the present driven-system framework, while its Planck-scale substitution should be regarded only as a dimensional limiting estimate.

**5. Universal Operational Criterion $\mathcal{A}_n$**

For a general driven Hamiltonian $H(\lambda(t))$, the natural normalization scale is the instantaneous spectral gap $\Delta_n(t) = \min_{m\neq n}|E_m - E_n|$. Here $\Delta_n(t)$ denotes the instantaneous minimal spectral separation between the reference state $|n\rangle$ and all other instantaneous eigenstates of the driven Hamiltonian. Near exact degeneracies, $\Delta_n \rightarrow 0$ and the instability parameter diverges, signaling the breakdown of adiabatic following. The criterion is therefore well-defined away from exact spectral degeneracies, where the local adiabatic manifold remains differentiable. In units $\hbar = 1$, we define the universal geometric instability parameter:

$$\mathcal{A}_n(t) \equiv \frac{v_{FS}{}^{(n)}(t)}{\Delta_n(t)} = \frac{\sqrt{g_{\mu\nu}\dot{\lambda}^{\mu}\dot{\lambda}^{\nu}}}{\Delta_n(t)} \text{ . (19)}$$

This is the central object of the present framework: a dimensionless, gauge-invariant, and basis-independent local measure of non-adiabaticity. The criterion for local instability is:

$$\mathcal{A}_n \ll 1\text{: } \textit{adiabatic} \quad | \quad \mathcal{A}_n \sim 1\text{: } \textit{onset of instability} \text{ . (20)}$$

The geometric criterion can also be expressed directly through the instantaneous non-adiabatic transition amplitudes. For the instantaneous eigenstates $|n(\lambda(t))\rangle$ of the driven Hamiltonian $H(\lambda(t))$, one obtains:

$$\langle m|\partial_t n\rangle = \frac{\langle m|\partial_t H|n\rangle}{E_n - E_m},\ m \neq n. \text{ (21)}$$

$$g_{\mu\nu}\dot{\lambda}^{\mu}\dot{\lambda}^{\nu} = \Sigma_{m\neq n}|\langle m|\partial_t n\rangle|^2 = v_{FS}{}^2 \text{ . (22)}$$

$$\mathcal{A}_n = \sqrt{\Sigma_{m\neq n}|\langle m|\partial_t n\rangle|^2}/\Delta_n \text{ . (23)}$$

This representation establishes the direct connection between the Fubini–Study evolution speed and the instantaneous transition amplitudes responsible for adiabatic breakdown. For the parametrically driven oscillator with $\Delta_0=\Omega$ and single control parameter $\lambda=\Omega(t)$, the general criterion reduces to the previously introduced AMT parameter:

Using the spectral representation of the QGT:

$$g_{\mu\nu}^{(n)} = Re\, \Sigma_{m\neq n} \frac{\langle n|\partial_\mu H|m\rangle\langle m|\partial_\nu H|n\rangle}{(E_m - E_n)^2}\ . \quad (24)$$

substituting into Eq. (19) gives $\mathcal{A}_n$ as a sum of transition-weighted contributions, each with enhanced inverse-gap sensitivity near spectral degeneracies and quantum critical points. This explains the divergence tendency of $\mathcal{A}_n$ near quantum critical points where the spectral gap closes. In particular, $\mathcal{A}_n \to \infty$ at quantum phase transitions where $E_m - E_n \to 0$, providing a geometric interpretation of the onset of Kibble–Zurek-type dynamics [13,14].

Near a quantum critical point $\lambda_c$, the spectral gap closes with a characteristic power law:

$$\Delta_n(t) \sim |\lambda - \lambda_c|^{z\nu}\ , \quad (25)$$

where z is the dynamical critical exponent and ν is the correlation-length exponent. Substituting into the definition of $\mathcal{A}_n$:

$$\mathcal{A}_n \sim \frac{v_{FS}}{|\lambda-\lambda_c|^{z\nu}} \ \to\ \infty \quad (26)$$

The condition $\mathcal{A}_n \sim 1$ defines the "freeze-out" moment $t^*$, the instant at which the geometric speed of the driven state equals the instantaneous spectral accessibility:

$$\mathcal{A}_n(t^*) = 1 \ \Rightarrow\ |\lambda(t^*) - \lambda_c|^{z\nu} = v_{FS}(t^*)\ . \quad (27)$$

This is a local, operational form of the Kibble–Zurek criterion [13,14]. The standard KZM requires integration along the full protocol to determine the defect density. The geometric criterion $\mathcal{A}_n \sim 1$ identifies the freeze-out point $t^*$ from local information alone: the instantaneous FS speed and the instantaneous gap. No knowledge of the future trajectory is required. For a linear ramp $\lambda(t)=\lambda_c+Yt$ with velocity $Y$ and the finite $v_{FS}$ near the freeze-out point, the freeze-out time scales as $t^* \sim |Y|^{-\nu z/(1+\nu z)}$, recovering the standard KZM scaling from the local geometric threshold.

Physical interpretation: the divergence of $\mathcal{A}_n$ at a quantum phase transition is not a failure of the criterion but its most important prediction. It signals that the geometric speed of the driven state diverges relative to the instantaneous spectral accessibility — the system can no longer follow adiabatically in the thermodynamic limit as the gap closes. This gives the Kibble–Zurek freeze-out a direct geometric interpretation: it is the moment when the Fubini–Study speed saturates the spectral gap.

The AMT parameter $\eta$ is not replaced by $\mathcal{A}_n$; rather, it is recovered as the oscillator-specific normalization of the general criterion. For the parametric oscillator with gap $\Delta_0 = \Omega$ and single parameter $\lambda = \Omega$ (in units $\hbar = 1$):

$$\mathcal{A}_0 = \frac{\sqrt{g_{tt}}}{\Omega} = \frac{\eta\Omega/\sqrt{8}}{\Omega} = \frac{\eta}{\sqrt{8}} \,. \quad (28)$$

confirming that $\mathcal{A}_0 \sim 1$ coincides with $\eta \sim \sqrt{8}$. The operational AMT threshold $\eta \approx 1$ introduced in our previous studies [6, 7] should therefore be interpreted as an order-of-magnitude crossover criterion rather than an exact universal constant. The present geometric normalization identifies the properly normalized instability parameter $\mathcal{A}_0$, yielding the exact single-mode relation $\mathcal{A}_0 = \eta/\sqrt{8}$. Consequently, the universal geometric instability threshold $\mathcal{A}_0 = 1$ corresponds to $\eta = \sqrt{8}$ in the parametrically driven oscillator realization, while the earlier criterion $\eta = O(1)$ remains a practically useful indicator of the crossover from adiabatic to non-adiabatic dynamics.

In the oscillator realization, nonlinear self-limitation is conveniently characterized by $\xi = \eta/U$. In a general quantum-geometric formulation, the corresponding stabilization criterion is expressed through the suppression of the normalized geometric parameter: $\mathcal{A}_n^{(U)} < \mathcal{A}_n$. The three levels of the hierarchy are therefore: (i) $\mathcal{A}_n$ — universal, any $H(\lambda)$; (ii) $\eta = \sqrt{8}\mathcal{A}_0$ — oscillator realization; (iii) $\xi = \eta/U$ — oscillator with nonlinear regulator.

**6. Nonlinear Metric Deformation by *U*.**

We consider the Hamiltonian with occupation-dependent nonlinear regulator [6]:

$$\hat{H}(t) = \Omega(t)a^{\dagger}a + U(a^{\dagger}a)^2 + G(t)(a^{\dagger 2} + a^2) \quad (29)$$

In the mean-field approximation, U induces an occupation-dependent frequency shift:

$$\Omega_{eff}(t) = \Omega(t) + U\langle n\rangle \ . (30)$$

where $\Omega_{eff}(t)$ is the nonlinear occupation-renormalized instantaneous spectral scale. For the stabilizing regime considered throughout this work, $U > 0$. Substituting into the FS metric formula:

$$g_{tt}^{(U)} = \frac{\dot{\Omega}^2}{8\Omega_{eff}{}^2} \approx \frac{\eta^2\Omega^2}{8} \cdot \frac{1}{(1 + U\langle n\rangle/\Omega)^2} \ . (31)$$

Since $\langle n\rangle \geq 0$, we have $g_{tt}^{(U)} \leq g_{tt}$ for all $t$. The quantum metric is compressed by the nonlinear regulator. The effective dimensionless speed becomes:

$$\eta_{eff} = \frac{|\dot{\Omega}|}{\Omega_{eff}{}^2} < \eta \ . \ (32)$$

This is a self-limiting mechanism: as occupation grows, $\eta_{eff}$ decreases, reducing the geometric speed and preventing parametric divergence. The crossover parameter:

$$\xi = \frac{\eta}{U} = \frac{geometric\ speed}{metric\ compression\ strength} \ . (33)$$

provides the compact geometric criterion: bounded dynamics require $\xi < \xi_0 \approx 1/4$ [6].

Beyond the mean-field replacement $\Omega_{eff} = \Omega + U\langle\hat{n}\rangle$, one may define the operator-valued effective spectral scale $\hat{\Omega}_{eff} = \Omega + U\hat{n}$. Its expectation value gives the mean-field result, whereas its second moment gives the rms effective spectral scale:

$$\Omega_{eff,rms}^2 = \langle\hat{\Omega}_{eff}^2\rangle = (\Omega + U\langle\hat{n}\rangle)^2 + U^2\, Var(\hat{n}), \ (34)$$

where $\mathrm{Var}(\hat{n}) = \langle\hat{n}^2\rangle - \langle\hat{n}\rangle^2$ is the occupation-number variance (the variance of the photon/magnon number). Thus, the variance term is not an ad hoc correction but follows directly from the second moment of the operator-valued spectral scale. The variance contribution in Eq. (34) follows directly from the second moment of the operator-valued spectral scale and therefore represents the exact contribution of occupation-number fluctuations to the rms effective frequency. The present correction is exact for the operator-valued spectral scale $\Omega_{\mathrm{eff}} = \Omega + U\hat{n}$ and captures the leading effect of occupation-number fluctuations through the second moment $\mathrm{Var}(\hat{n})$. In strongly correlated, topological, or highly non-Gaussian states, higher-order cumulants may generate additional corrections beyond the variance contribution considered here. In this sense, Eq. (34) provides a systematic beyond-mean-field extension of the average nonlinear shift. While the leading mean-field approximation renormalizes the effective spectral scale through the average occupation, finite occupation fluctuations are expected to generate additional broadening of the instantaneous excitation spectrum. The present form therefore provides a controlled beyond-mean-field extension of the mean-field description while retaining the exact second-moment contribution of occupation-number fluctuations. The corrected metric becomes:

$$g_{tt}^{(U,full)} = \frac{\eta\Omega^2}{8} \cdot \frac{1}{(1 + U\langle n\rangle/\Omega)^2 + (U/\Omega)^2 \cdot Var(n)} \;. \quad (35)$$

In the strong-blockade regime $U \gg \eta\Omega$, the system occupies a near-Fock state with sub-Poissonian statistics: $\mathrm{Var}(\hat{n}) \rightarrow 0$. The mean-field result Eq. (31) is therefore self-consistently accurate in the very regime where geometric stabilization is strongest. At intermediate coupling, the fluctuation term $U^2 \cdot \mathrm{Var}(\hat{n})$ provides an additional positive contribution to $\Omega_{eff}$, further suppressing $g_{tt}^{(U,full)}$ below the mean-field estimate. This connection places the fluctuation correction within the broader context of quantum geometry in interacting systems [15]. This means the mean-field approximation is a conservative lower bound on the metric compression: quantum fluctuations reinforce, rather than undermine, the self-limiting mechanism .

The variance correction represents the leading beyond-mean-field contribution associated with occupation-number fluctuations. For Gaussian, thermal, coherent, and weakly interacting states, the second cumulant provides the dominant correction to the effective spectral scale. In strongly correlated, critical, or topological many-body phases, higher-order cumulants may become quantitatively important and generate additional corrections beyond Eq. (35). The present treatment should therefore be regarded as a controlled leading-order fluctuation expansion rather than a complete many-body resummation.

The mean-field approximation remains quantitatively reliable when occupation-number fluctuations remain small compared to the average nonlinear shift, i.e. Var() $\ll \langle\hat{n}\rangle^2$, or equivalently, $U\hat{n}) \ll \Omega + U\langle n\rangle$. In this regime, the fluctuation-induced correction of Eq. (34) provides only a perturbative renormalization of the effective spectral scale, and the instability criterion $\mathcal{A}_n$ remains governed by the mean-field geometry. For coherent states, Var($\hat{n}$)=$\langle\hat{n}\rangle$, implying a relative mean-field error that scales as $1/\sqrt{\langle\hat{n}\rangle}$ and therefore vanishes in the large-occupation limit. Strongly correlated regimes with Var($\hat{n}$)$\approx\langle\hat{n}\rangle^2$ may require a full many-body evaluation of the quantum geometric tensor beyond the present approximation.

In the adiabatic limit $A_n \ll 1$, the present framework reduces to the standard quantum-geometric description of adiabatic transport. The real part of the QGT controls the local metric susceptibility to non-adiabatic excitation, while the imaginary antisymmetric part gives the Berry curvature and hence the geometric phase accumulated during cyclic evolution. Thus, the proposed criterion does not replace Berry/Aharonov–Anandan geometry, but completes the same QGT structure by identifying the metric sector as the local non-adiabatic channel

## 7. Many-Mode Quantum-Geometric Instability Matrix

The scalar criterion $\mathcal{A}_n$ captures non-adiabaticity in a single parameter direction. In real physical systems, magnon modes, multimode photonic cavities, or many-qubit processors, several parameters vary simultaneously and their collective non-adiabatic response is not captured by individual single-mode parameters. The key physical question is: which collective direction in parameter space drives the system most rapidly toward instability? We answer this by promoting $\mathcal{A}_n$ to a matrix.

Mathematically, the multimode instability criterion naturally acquires the structure of a quadratic form in parameter-velocity space:

$$\mathcal{A}_n^2 \equiv \frac{\dot{\lambda}^T g^n \dot{\lambda}}{\Delta_n^2} \quad (36).$$

Eq. (36) can be written compactly as $\mathcal{A}_n^2 = \dot{\lambda}_{\,i}\mathcal{A}_{ij}^{(n)}\,\dot{\lambda}_{\,j}$ . For $H(\lambda^1,...,\lambda^N)$, the quantum-geometric instability matrix is:

$$\mathcal{A}_{ij}^{(n)} \equiv \frac{g^n}{\Delta_n^2} \quad (37).$$

The maximal collective instability channel is determined by the largest eigenvalue of the normalized geometric tensor:

This $N \times N$ symmetric positive-semidefinite matrix captures both single-channel non-adiabaticity (diagonal elements) and geometric cross-correlations between parameter channels (off-diagonal elements). The stability criterion generalizes naturally to a spectral condition:

$$\lambda_{max}\left(\mathcal{A}_{ij}^{(n)}\right) < \mathcal{A}_{crit}^2 \quad . (38)$$

where $\lambda_{max}\left(\mathcal{A}_{ij}^{(n)}\right)$ denotes the largest eigenvalue of the normalized instability matrix. The eigenvectors of $\mathcal{A}_{ij}$ identify the dominant collective instability channels, the directions in parameter space along which the system is most sensitive to non-adiabatic excitation. The corresponding eigenvalues quantify how close each channel is to the instability threshold. Channels with eigenvalues near $\mathcal{A}_{crit}$ are geometrically fragile; channels with small eigenvalues are robust.

Physical interpretation: the instability matrix $\mathcal{A}_{ij}$ is the multimode analogue of the adiabaticity condition. It reveals that in coupled systems, instability is not necessarily triggered by the most rapidly varying individual mode, but by the collective direction of maximum geometric sensitivity — which may combine slow individual drives into a fast collective motion in projective Hilbert space.

When each mode i has its own frequency $\Omega_i$(t) as an independent parameter ($\lambda$=( $\Omega_1$,..., $\Omega_N$) independent mode frequencies, the instability matrix takes the form:

$$\mathcal{A}_{ij}^{(n)} = \frac{g_{ij}{}^{(n)}}{\Omega_i \cdot \Omega_j}, \quad \mathcal{A}_{ii} = \frac{\eta_i^2}{8} \quad . (39)$$

Here $g_{ij}$ denotes the multimode Fubini–Study metric tensor in the space of independently driven mode frequencies. For the normalization adopted throughout this work, $\mathcal{A}_{crit}$=1. The diagonal elements $\mathcal{A}_{ii} = \frac{\eta_i^2}{8}$ recover the single-mode scalar criterion for each mode independently. The off-diagonal elements $\mathcal{A}_{ij}$ measure the geometric coupling between modes i and j induced by the structure of the quantum state manifold. When modes are driven independently and their vacuum states factorize, $g_{ij} = 0$ and the matrix is diagonal; the stability criterion then reduces to $max_i(\frac{\eta_i^2}{8}) < \mathcal{A}_{crit}$, which is a direct multimode generalization of $\eta$<1.

When single external drive parameter controlling all modes yields a scalar $\mathcal{A}_n$ with summed modal contributions and is recovered as a special case of Eq. (29).

The spectral structure of $\mathcal{A}_{ij}$ thus provides a geometric classification of multimode non-adiabatic dynamics: it separates robust (small eigenvalue) from fragile (large eigenvalue) collective modes and identifies the instability-dominant subspace of parameter space without requiring mode-by-mode analysis. Related geometric tensor structures in many-body systems and collective quantum response have been studied in the context of fidelity susceptibility and non-Abelian quantum geometry [16,17].

For large many-body Hilbert spaces, the explicit construction of the instability matrix $\mathcal{A}_{ij}$ may become computationally demanding due to the exponential growth of the quantum-state manifold dimension. In such regimes, practical applications will likely require reduced-order descriptions, tensor-network approximations, or phenomenological coarse-grained geometric models.

In large correlated many-body systems, the maximal eigenvalue of the instability matrix is expected to acquire a characteristic scaling form:

$$\lambda_{max}(\mathcal{A}_{ij}) \sim N^{\alpha} \text{ , (40)}$$

where $N$ is the effective number of coupled degrees of freedom participating in the driven dynamics and the exponent $\alpha$ depends on the correlation structure of the quantum state manifold. Independent modes correspond to weak or subextensive scaling, whereas strongly correlated or critical states may generate enhanced collective geometric susceptibility through superextensive growth of the dominant instability channel.

The present framework therefore provides the formal geometric structure, while scalable many-body implementations remain an important open direction. Importantly, practical instability detection does not require reconstruction of the full matrix $\mathcal{A}_{ij}$. In many applications, only the dominant eigenvalue $\lambda_{max}$ and its associated collective mode are required. These quantities can often be estimated directly through iterative eigensolvers, tensor-network contractions, or variational subspace methods, reducing the computational cost from full matrix reconstruction to extraction of the leading instability channel.

To illustrate the matrix formalism concretely, consider two coupled bosonic modes:

$$\hat{H} = \hat{p}_1^2/2 + V_{11}\hat{x}_2^2/2 + \hat{p}_2^2/2 + V_{22}\hat{x}_2^2/2 + J\hat{x}_1\hat{x}_2 \quad (41)$$

where $J$ is the inter-mode coupling. For the two-mode vacuum state, the FS metric tensor has diagonal elements from single-mode contributions and off-diagonal elements from J-induced correlations:

$$g_{11} = \frac{1}{8\Omega_1^2}, \; g_{12} \sim \frac{3J^2}{8(\Omega_1^2 - \Omega_2^2)^2} \quad (42)$$

The quantum-geometric instability matrix Eq. (37) is a 2×2 symmetric matrix:

$$\mathcal{A}_{ij} = \begin{pmatrix} \mathcal{A}_{11} & \mathcal{A}_{12} \\ \mathcal{A}_{12} & \mathcal{A}_{22} \end{pmatrix} \quad (43)$$

Its eigenvalues are:

$$\lambda_{\pm} = \frac{\mathcal{A}_{11} + \mathcal{A}_{22}}{2} \pm \sqrt{\frac{(\mathcal{A}_{11} - \mathcal{A}_{22})^2}{4} + {\mathcal{A}_{12}}^2} \quad (44)$$

The large eigenvalue $\lambda_{max} \equiv \lambda_+$ identifies the dominant instability channel. When $J = 0$ (decoupled modes), $\mathcal{A}_{12}$=0 and $\lambda_{max}$= max($\mathcal{A}_{11}$, $\mathcal{A}_{22}$) — the scalar criterion suffices. When J ≠ 0, the coupling hybridizes the two modes and the dominant instability channel is a linear combination of both mode frequencies. Crucially, $\lambda_{max}$>max($\mathcal{A}_{11}$, $\mathcal{A}_{22}$)): the coupled system is more unstable than either individual mode. This collective enhancement of instability is invisible to the scalar criterion and is the key physical content of the matrix formalism.

Physical interpretation: inter-mode coupling hybridizes the instability channels. The dominant eigenvector of $\mathcal{A}_{ij}$ points in the direction of maximum collective geometric sensitivity , a superposition of both mode frequencies that can be driven to instability even when each mode individually satisfies $\eta_i < 1$. This is a genuinely many-mode effect with no single-mode analogue.

**8. Theorem (Nonlinear Geometric Compression)**

The mean-field result of Section 6 showed that the nonlinear regulator *U* compresses the FS metric. But mean-field assumes a specific occupation statistics. The physical question here is deeper: is metric compression an intrinsic geometric consequence of *U*, or does it depend on the quantum state of the system? We show that it is intrinsic — a direct consequence of the monotonic spectral shift induced by *U*, independent of the occupation distribution.

For the nonlinear Hamiltonian (22), the *U*-renormalized QGT through the spectral representation is:

$$g_{\mu\nu}^{(n,U)} = Re\, \Sigma_{\mathrm{m}} \neq {}_{\mathrm{n}} \frac{\langle n|\partial_\mu H|m\rangle\langle m|\partial_\nu H|n\rangle}{(E_{\mathrm{m}}-E_{\mathrm{n}}+U(f(m)-f(n))^2} . \quad (45)$$

Assume that the occupation-basis eigenstates are ordered by increasing occupation number and that *f*(n) is a monotonically increasing function:

$$H_U = H_0 + Uf(\hat{n}). \, (46)$$

For eigenstates |m⟩ of the occupation basis, the nonlinear spectrum is:

$$E_m^{(U)} = E_m^{(0)} + U\, f(m). \, (47)$$

Accordingly, the renormalized spectral difference is:

$$\Delta_{mn}^{(U)} = E_m^{(U)} - E_n^{(U)} = \Delta_{mn}^{(0)} + U[f(m)\text{-}\, f(n)]. \, (48)$$

Since *f(m)-f(n)*≥0 for m>n and *U*>0, the nonlinear regulator increases the spectral gaps $|\Delta_{mn}^{(U)}| \geq |\Delta_{mn}^{(0)}|$ and therefore $(\Delta_{mn}^{(U)})^2 \geq (\Delta_{mn}^{(0)})^2$. Since all squared spectral denominators are strictly positive, inversion gives $1/(\Delta_{mn}^{(U)})^2 \leq 1/(\Delta_{mn}^{(0)})^2$ provided the transition matrix elements of $\partial_\mu H$ remain bounded under the nonlinear deformation. Therefore, each denominator satisfies $|E_m^{(U)} - E_n^{(U)}| \geq |E_m^{(0)} - E_n^{(0)}|$ the inequality holds term by term:

$$g_{\mu\nu}^{(n,U)} \leq g_{\mu\nu}^{(0)} \Rightarrow \mathcal{A}_n^{(U)} \leq \mathcal{A}_n \quad . (49)$$

The inequality is understood element-wise in the spectral representation. Using Eq. (34), one obtains the bound Eq. 49, which shows that for *U*>0 the nonlinear regulator geometrically compresses the quantum metric. The mean-field result of Section 6, including the beyond-mean-field correction via Var(n), are special cases of this general inequality. Eq. (49) is the geometric foundation for the self-limitation theorem of Section 9.

Physical interpretation: nonlinear interactions renormalize the geometry of the quantum state manifold. Interaction-induced renormalization of quantum geometry has recently emerged as an important theme in strongly correlated quantum systems and flat-band quantum matter [15]. The regulator U effectively increases the curvature of the instantaneous energy spectrum, making the state space locally "stiffer" — harder to traverse at speed. This is a purely geometric effect: *U* does not damp the dynamics, it deforms the metric through which the dynamics propagate.

## 9. Self-Limited Non-Adiabatic Dynamics

The geometric compression of Section 8 shows that *U* reduces $\mathcal{A}_n$ at every instant. But does a reduced instantaneous geometric speed guarantee that the trajectory remains globally bounded? In principle, a system could accumulate small excursions that diverge over time. Assume that the accessible low-occupancy manifold is compact with respect to the induced Fubini–Study metric. Furthermore, assume that the occupation functional N(t) is continuous on this manifold. Then bounded Fubini–Study trajectories correspond to bounded occupation excursions within the accessible region of the state manifold. We show that this does not happen: bounded geometric speed on the state manifold implies bounded arc length and hence bounded occupancy.

Theorem (Geometric Self-Limitation): Let $H(\lambda(t))$ be a driven quantum Hamiltonian with occupation-dependent nonlinear regulator $U > 0$ satisfying the monotonic gap condition of Section 8. Then for any initial state in the low-occupancy manifold

$$\mathcal{A}_n^{(U)}(t) < \mathcal{A}_n(t) \quad . (50)$$

Combined with the compactness and growth assumptions below, this implies bounded trajectories. A more rigorous formulation can be developed through an occupation functional:

$$N(t)=\langle\hat{n}\rangle \; (51).$$

$$\frac{dN}{dt} \le C\mathcal{A}_n^{(U)}(t) - \Gamma_U(N) \; . (52).$$

where $\Gamma_U(N)$is the nonlinear geometric suppression functional generated by the $U$-induced metric compression. Where $C$>0 is a Lipschitz constant relating occupation growth to the geometric speed on the low-occupancy manifold. Assume that the geometric driving contribution is bounded

$$C\mathcal{A}_n^{(U)}(t) \le M < \infty, \; (53)$$

and that the nonlinear suppression functional satisfies $\Gamma_U(N) \to \infty$. Then there exists a finite N* such that $\Gamma_U(N)$>$M$ for all $N$>$N$*. Therefore,

$$\frac{dN}{dt} \le M - \Gamma_U(N) < 0, \; N > N^*, \; (54)$$

which implies

$$\exists N^* : \frac{dN}{dt} < 0 \quad \forall N > N^* \; (55).$$

Equivalently, the occupation can be bounded directly by the Fubini–Study arc length. Since $N$(t) is assumed continuous on the accessible low-occupancy manifold, and since the state trajectory has finite Fubini–Study length $L_{FS}(t)$:

$$L_{FS}(t) = \int_0^t v_{FS}(t')dt' \quad (56)$$

there exists a monotone function $F$ such that

$$N(t) \leq F(L_{FS}(t)) \quad (57).$$

For finite $L_{FS}(t)$, the occupation therefore remains bounded. This provides the geometric form of the self-limitation mechanism: bounded distinguishability length implies bounded occupation excursion within the compact accessible manifold.

Thus, the occupation cannot grow beyond the finite absorbing bound $N^*$. This provides the Lyapunov-type confinement mechanism underlying geometric self-limitation. The geometric interpretation of constrained quantum evolution is broadly consistent with earlier geometric formulations of quantum mechanics and stability theory [18, 19].

Proof sketch. By Eq. (49), $\mathcal{A}_n^{(U)} < \mathcal{A}_n$ at every t, so the geometric speed of the driven state is strictly reduced. Simultaneously, the spectral gap $\Delta_n$ increases monotonically with U, further reducing $\mathcal{A}_n^{(U)}$. Therefore $\mathcal{A}_n^{(U)} < \mathcal{A}_n$ at every instant. Integrating over time, the total arc length growth is suppressed relative to the unregulated dynamics. For the compact low-occupancy region considered here, bounded Fubini–Study arc length restricts the accessible subset of the state manifold and therefore limits occupation growth.

The oscillator realization $\xi = \eta/U < 1/4$ from [6] is a special case of Eq. (39) in which $\mathcal{A}_n^{(U)}$ is expressed through the effective AMT parameter $\eta_{\text{eff}}$. The theorem applies to driven quantum systems admitting a monotonic occupation-dependent spectral regulator.

Physical interpretation: the self-limitation theorem establishes that nonlinear interactions do not merely slow the system down — they geometrically confine it. The driven state is not suppressed by dissipation or damping, but by the curvature of the quantum state manifold itself: the occupied region of projective Hilbert space becomes geometrically inaccessible under strong driving precisely because the nonlinear regulator deforms the metric in a way that forbids fast traversal.

## 10. Outlook and Extensions

The framework of Sections 5–9 opens three directions that extend the geometric non-adiabaticity criterion into broader domains of physics. Each direction connects $\mathcal{A}_n$ to an established field through a natural bridge, while introducing new physical content.

### 10.1. Quantum Thermodynamics: Irreversible Work and Entropy Production

In finite-time quantum thermodynamics, driving a system non-adiabatically generates irreversible work — energy dissipated as quantum excitations above the instantaneous ground state $W_{irr}$.This scaling follows from the transition-amplitude representation of Section 5. To leading order, the instantaneous excitation rate is controlled by

$$P_{ex} \sim \Sigma_{m \neq n} |\langle m|\partial_t n\rangle|^2 \;. (58)$$

Using the spectral representation of the quantum geometric tensor,

$$g_{tt}^{(n)} = \Sigma_{m \neq n} |\langle m|\partial_t n\rangle|^2 \quad . (59)$$

Using the definition of $\mathcal{A}_n$, this gives

$$\Sigma_{m \neq n} |\langle m|\partial_t n\rangle|^2 = \Delta_n^2 \mathcal{A}_n^2 \;. (60)$$

If the characteristic excitation energy is $\Delta_n$, the local irreversible-work rate scales as

$$W_{irr} \sim \int \Delta_n^3 \mathcal{A}_n^2(t) dt \;. \; (61)$$

The relation provides that $\mathcal{A}_n$ can be served as a local measure of this irreversibility.

This expression states that the local rate of irreversible work is proportional to $\mathcal{A}_n^2$(t) the squared geometric instability parameter — weighted by the instantaneous spectral gap. The physical content is direct: the faster the system traverses its state manifold relative to the spectral gap (large $\mathcal{A}_n$), the more irreversible excitations are produced per unit of parameter change.

A rigorous derivation of Eq. (19) requires extending the present framework to open quantum systems, where the Fubini–Study metric is replaced by the Bures or quantum Fisher metric for mixed states [16]. This extension is straightforward in principle: the quantum Fisher information metric $g_{\mu\nu}^{QF}$ for a density matrix $\rho(\lambda)$ satisfies the same spectral representation as Eq. (21), with eigenstate overlaps replaced by mixed-state fidelity derivatives. The parameter $\mathcal{A}_n$ then becomes a measure of local fidelity susceptibility, directly accessible in experiments through dynamical decoupling protocols on quantum processors [6]. Connections between geometry, irreversibility, and finite-time quantum thermodynamics have also been explored in broader contexts of quantum thermodynamic geometry and quantum information thermodynamics [20–22].

The present formulation is rigorously established for pure-state unitary dynamics. While the extension to mixed states through the Bures or quantum Fisher metric is conceptually natural, the full influence of decoherence, dissipation, and environment-induced metric renormalization on the local instability parameter $\mathcal{A}_n$ remains to be systematically investigated. In open quantum systems, geometric non-adiabaticity may compete with relaxation and dephasing processes, potentially modifying both the instability threshold and the self-limitation mechanism.

Physical interpretation: $\mathcal{A}_n$ is the local geometric cost of non-adiabatic driving. Minimizing $\mathcal{A}_n(t)$ along a protocol tends to minimize local entropy production and irreversible work. This provides a new geometric principle for optimal quantum control: this suggests that geodesic trajectories of the quantum-state manifold may provide thermodynamically efficient driving protocols.

## 10.2. Berry Curvature, Topology, and Geometric Protection

The present framework operates in the real Gaussian gauge where the Berry connection vanishes (Eq. 7) and the QGT reduces to its real symmetric part. The physical question is: what happens when the driving protocol has a complex phase structure that activates the imaginary part of the QGT?

For the coherent-state example introduced above, one obtains:

$$\Omega_{\Omega\theta} = 2\, Im\langle\partial_\Omega\psi_0|\partial_\theta\psi_0\rangle = 2|\alpha| \text{ . } (62)$$

For a coherent state with complex amplitude $\alpha = |\alpha|e^{i\theta}$, this evaluates to $\Omega_{\Omega\theta} = 2|\alpha|$, which is nonzero and generates a Berry phase upon cyclic evolution in ($\Omega$, $\theta$)-space. The FS metric $g_{\mu\nu}$ remains gauge-invariant and unaffected. However, the coupling between the metric sector $\mathcal{A}_n$ and the curvature sector $\Omega_{\mu\nu}$ under nonlinear driving may generate topologically protected bounded regimes — regions of parameter space where both geometric speed and Berry curvature simultaneously enforce stability. The corresponding Chern number would characterize the topology of the curvature sector associated with the driving protocol. A closed two-parameter driving loop C in the ($\Omega$,$\theta$) manifold accumulates the geometric phase

$$\gamma_B = \oint C \mathcal{A}_\mu \mathrm{d}\,\lambda_\mu = \frac{1}{2}\int_S \ \Omega_{\mu\nu}\,\mathrm{d}S_{\mu\nu} \ \ (63).$$

If the Berry-curvature sector generates a finite geometric phase while the metric sector simultaneously satisfies

$$\mathcal{A}_n(\lambda(t)) < \ \mathcal{A}_{\mathrm{crit}} \ ,(64)$$

along the same closed trajectory, the evolution remains geometrically bounded over one driving cycle. Where $\mathcal{A}_{\mathrm{crit}}$ is the local instability threshold introduced in Eq. (20). This defines a proposed coupled metric-curvature stabilization regime.

A quantitative formulation of this condition will require comparing the curvature-induced transport scale with the local instability scale.

$$|\Omega_{\mu\nu}\dot{\lambda}^\mu\dot{\lambda}^\nu| < \Delta_n^2\mathcal{A}_{crit}^2 \ \ . \ (65)$$

We propose the dimensionless curvature-instability ratio. A natural local measure of coupled topological stabilization is provided by the curvature-instability scalar $\mathcal{T}$:

$$\mathcal{T}(\lambda) = \sqrt{\Omega_{\mu\nu}\Omega^{\mu\nu}}\,\mathcal{A}_n^{-1} \ , (66)$$

which compares the local Berry-curvature strength to the geometric instability growth rate. Large values of $\mathcal{T}$ correspond to regimes where topological transport dominates over metric instability generation.

In this sense, nonlinear geometric stabilization may acquire a partially topological character: sufficiently strong Berry-curvature sectors may dynamically suppress unstable metric directions on the quantum-state manifold.

Physical interpretation: the full QGT unifies two apparently separate phenomena — non-adiabatic instability (metric sector) and geometric phase (curvature sector). In the two-parameter case, these interact: a system that is metrically unstable $\mathcal{A}_n \sim 1$ may still be topologically protected if the Berry curvature creates a geometric barrier in parameter space. This suggests a possible topological self-limitation mechanism. More broadly, the interplay between Berry curvature, quantum geometry, and topological response has become a central theme in modern quantum matter and geometric quantum dynamics [23,24].

**10.3. Inverse theorem: metric compression bounds Berry curvature.**

The nonlinear compression theorem has a topological corollary. To make the connection with the curvature sector explicit, we consider the standard squeezed-vacuum representation of the parametrically driven oscillator. The instantaneous quantum state is parameterized by the squeezing amplitude $r$ and the squeezing phase $\varphi$, which provide natural coordinates on the corresponding quantum-state manifold. In this representation, both the Fubini–Study metric and the Berry curvature admit closed analytical expressions.

For the squeezed-vacuum realization of the driven system, the exact results of Section 10.2 give the Berry curvature $|\Omega_{r\varphi}|=\sinh(2r)/2$ and the metric $g_{\varphi\varphi}= \sinh^2(r)$. The compression theorem (Eq.(43): $g^{(U)} \leq g$) implies that $U$ reduces the effective squeezing parameter: $r(U) \leq r(0)$. Since $\sinh(2r)$ is monotonically increasing, this immediately gives:

$$|\Omega_{r\varphi}^{(U)}| = |\Omega_{r\varphi}^{(0)}| \cdot \frac{sinh(2r(U))}{sinh(2r(0))} \leq |\Omega_{r\varphi}^{(0)}| \quad . (67)$$

The nonlinear regulator $U$ not only compresses the quantum metric but simultaneously suppresses the Berry curvature. The self-limited system is therefore also topologically stabilized: the Berry phase accumulated per driving cycle is bounded from above by the Berry phase of the $U = 0$ system.

More precisely: the ratio $||\Omega_{r\varphi}^{(U)}|/|\Omega_{r\varphi}^{(0)}|$ = sinh(2r($U$))/sinh(2r(0)) ≤ 1 is controlled by the same metric compression factor that bounds the instability parameter $\mathcal{A}_n^{(U)} \leq \mathcal{A}_n$. Metric stabilization and topological suppression are therefore two aspects of the same geometric compression mechanism.

Physical interpretation: the nonlinear regulator $U$ acts as a joint geometric and topological stabilizer. It simultaneously (i) reduces the FS metric speed, suppressing non-adiabatic excitations; (ii) reduces the Berry curvature, suppressing geometric phase accumulation. Both effects originate from the same nonlinear renormalization of the instantaneous spectral gaps, which geometrically compresses the quantum-state manifold. This stabilization mechanism is qualitatively different from conventional approaches based on dissipation or damping.

**10.4. Unruh-like Vacuum Excitation and the Geometric Speed Limit (Exact Berry curvature for the squeezed vacuum).**

The discussion below should be understood as a formal geometric analogy. The present framework does not derive the Unruh effect, nor does it imply equivalence between accelerated observers in quantum field theory and driven quantum systems. Rather, both systems involve distinguishability growth on a quantum-state manifold and therefore exhibit mathematically related geometric structures.

For the single-mode squeezed vacuum |r,φ⟩ = $S$(r,φ)|0⟩ with squeezing amplitude r and phase φ, the Berry connection and curvature are exactly calculable. The Berry connection in the φ direction is:

$$A_\varphi = i\langle r,\varphi|\partial_\varphi|r,\varphi\rangle = -\frac{sinh^2(r)}{2} \; . (68)$$

The Berry curvature in the (r,φ) plane is:

$$\Omega_{r\varphi} = \partial_r A_\varphi = -\frac{sinh(2r)}{2} \; . (69)$$

The quantum metric components are:

$$g_{rr} = 1, \;\; g_{\varphi\varphi} = sinh^2(r). (70)$$

These results are exact for all r ≥ 0 and φ ∈ [0, 2π). As r → ∞ (strong squeezing, near-threshold regime), $|\Omega_{r\varphi}|$ = sinh(2r)/2 ~ $e^{2r}/4$ → ∞ and $g_{\varphi\varphi}$ = sinh²(r) ~ $e^{2r}//4$ → ∞. Both the FS metric and the Berry curvature diverge simultaneously and at the same exponential rate. The Berry phase accumulated over one closed loop in (r,φ)-space at squeezing amplitude r is:

$$\gamma^{B} = \int_0^{2\pi} \int_0^{r} \Omega_{r\varphi}\, dr'\, d\varphi = \pi(cosh(2r) - 1)/2\ , (71)$$

which grows exponentially with *r*. The physical content is direct: as the squeezing amplitude approaches the regime where the geometric instability criterion $\mathcal{A}_n$ ~ 1 is activated, the accumulated Berry phase per driving cycle becomes large. Non-adiabatic instability and enhanced Berry-phase response therefore emerge simultaneously as the squeezing parameter increases, reflecting the common geometric origin of both effects in the quantum geometric tensor.

Physical interpretation: the QGT governs both phenomena. Its real part (FS metric $g_{\varphi\varphi}$) controls the local geometric speed and the instability threshold. Its imaginary part ($\Omega_{r\varphi}$) controls the Berry curvature and the accumulated geometric phase. Both quantities grow exponentially with the squeezing parameter *r*, reflecting the common geometric origin of the metric and curvature sectors of the QGT. This establishes a direct geometric connection between non-adiabatic breakdown and enhanced Berry-curvature response

Physical interpretation: the QGT governs both phenomena. Its real part (FS metric ) controls the local geometric speed and the instability threshold. Its imaginary part (controls the Berry curvature and the accumulated geometric phase. Both increase rapidly in the strong-squeezing regime and are controlled by the same underlying QGT structure.

The Unruh effect predicts that an observer accelerating through flat spacetime with acceleration *a* detects a thermal bath of Rindler particles at temperature $T = \hbar a/(2\pi k_B)$ [25]. The physical mechanism is that the accelerating observer's vacuum is distinguishable from the inertial vacuum, a statement directly about the geometry of the quantum state manifold. In both cases, the relevant quantity is the rate at which the overlap fidelity between neighboring vacuum states decreases under the evolution parameter. For infinitesimally close vacuum configurations, this distinguishability is quantified by the Fubini–Study line element

$$ds^2_{FS} = 1 - |\langle\psi(\lambda)|\psi(\lambda + d\lambda)\rangle|^2 \,. (72)$$

In our framework, a parametrically accelerated oscillator with $\mathcal{A}_n(t) \sim 1$ locally saturates the quantum speed limit: the instantaneous vacuum state becomes maximally distinguishable from the evolved state per unit of dimensionless time. The corresponding growth of the local Fubini–Study distance is therefore bounded by the same Mandelstam–Tamm geometric speed limit that controls unitary quantum evolution. The corresponding geometric acceleration of the quantum trajectory is naturally defined by:

$$a_{FS} = \frac{dv_{FS}}{dt}, \ (73)$$

which measures the local rate of change of distinguishability growth on the quantum-state manifold.

The existence of nonzero geometric acceleration implies that the quantum trajectory is locally non-geodesic on the projective Hilbert manifold.

The corresponding non-adiabatic excitation density is controlled by the local fidelity-loss rate.

$$\chi_{exc}(t) = 1 - |\langle\psi(t)|\psi(t + dt)\rangle|^2 \ (74).$$

For infinitesimal evolution,

$$\chi_{exc}(t) = ds^2_{FS} = v^2_{FS} dt^2_{FS} \ \ (75).$$

Using the definition of the normalized geometric instability parameter,

$$\chi_{exc}(t) = \Delta^2_n \mathcal{A}^2_n dt^2. (76)$$

Thus, the local excitation probability is controlled by $\mathcal{A}^2_n$ and becomes significant as $\mathcal{A}_n(t)$ approaches unity.

This establishes a geometric correspondence between local saturation of the quantum speed limit in driven systems and the distinguishability-based interpretation of the Unruh effect. The parameter $\mathcal{A}_0 \sim 1$ may therefore be interpreted as a local operational criterion for the onset of Unruh-like excitation behavior within the present geometric analogy.

At the level of dimensional scaling, the local excitation rate induced by geometric acceleration defines an effective temperature scale. Concretely, for a driven oscillator with $\Omega(t)$ varying at rate $\eta \sim 1$, the instantaneous vacuum evolves at the maximum geometric speed $v_{FS} = \Omega/(2\sqrt{2})$, producing Bogoliubov quasiparticles with a local excitation rate characterized by the geometric speed limit. This motivates the introduction of an effective excitation-temperature scale $T_{eff} \sim \eta\Omega/(2\pi k_B)$ as a dimensional measure of the local distinguishability-growth rate on the quantum-state manifold. The temperature scaling is therefore not postulated phenomenologically but emerges from the local fidelity-loss rate associated with accelerated traversal of the quantum-state manifold.

Consequently, the universal instability threshold $\mathcal{A}_0 \sim 1$ corresponds to $\eta \sim \sqrt{8}$ in the parametrically driven oscillator. The previously introduced criterion $\eta \sim 1$ should therefore be regarded as an order-of-magnitude crossover indicator rather than an exact universal threshold. In the nonlinear oscillator realization, the physically relevant control parameter becomes $\xi = \eta/U$, whose crossover value $\xi_0 \approx 1/4$ was established numerically in Ref. [6].

The nonlinear regulator $U$ arrests this divergence: above $\xi_0$, the metric compression of Section 8 bounds $T_{eff}$ at a finite value determined by the self-limitation theorem of Section 9. The boundedness of the effective excitation density follows from the self-limitation theorem of Section 9, which guarantees that the nonlinear regulator suppresses the growth of $\mathcal{A}_n$ beyond the instability crossover. The present argument does not derive a thermal density matrix. Rather, it defines an operational excitation-temperature scale through the local fidelity-loss rate. A genuine thermal spectrum would require an additional open-system or detector-response calculation. The broader relation between accelerated observers, vacuum distinguishability, and quantum fields in curved spacetime is discussed extensively in Refs. [25,26].

Within this interpretation, the nonlinear metric compression of Sections 8–9 acts as a geometric ultraviolet regulator: the local distinguishability growth rate is dynamically bounded by the nonlinear deformation of the quantum metric. In this sense, the nonlinear self-limitation mechanism dynamically restricts the accessible region of the quantum-state manifold, creating an effective geometric accessibility boundary for strongly driven states. The boundedness follows directly from the nonlinear geometric compression theorem of Section 8, which guarantees suppression of the local metric speed under monotonic spectral deformation.

In parametrically driven bosonic systems, the same geometric instability parameter controls the Bogoliubov pair-production amplitude through the non-adiabatic transition matrix elements of Section 5.

Physical interpretation: the self-limitation theorem of Section 9 has a thermodynamic reading in the Unruh-like regime: the nonlinear regulator $U$ acts as an effective geometric regulator that caps the excitation-temperature scale at a finite value. Within the present geometric framework, the metric deformation prevents the geometric instability parameter from diverging and therefore bounds the effective excitation-temperature scale. This suggests a geometric self-regulation mechanism ("geometric thermostating") arising from nonlinear metric compression.

## 11. Extension to Open Quantum Systems

The framework is formulated for closed (unitary) dynamics. Real systems couple to their environment through decoherence, relaxation, and thermal fluctuations. We extend the geometric instability criterion to open quantum systems described by the Lindblad master equation and identify the leading corrections to the instability threshold.

For a driven oscillator with energy decay rate $\gamma$:

$$\frac{d\rho}{dt} = -i[H,\rho] + \gamma\left(\hat{a}\rho\hat{a}^{\dagger}\dagger - \frac{1}{2}\{\hat{a}^{\dagger}\hat{a}\,,\rho\}\right) \quad (77)$$

For mixed states the FS metric is replaced by the Bures metric. For a Gaussian state with covariance matrix Σ it takes the same formal structure as Eq. (14):

$$g_{\mu\nu}^{Bures} = \frac{1}{4}T_r\left(\Sigma^{-1}\partial_{\mu}\Sigma \cdot \Sigma^{-1}\partial_{\nu}\Sigma\right)\ . (78)$$

Under weak driving at temperature $T$, the instantaneous state is a thermal Gaussian with covariance:

$$\Sigma(t) = (2n_{th}+1)\cdot\Sigma_{pure}(t), \quad n_{th} = \frac{1}{e^{\Omega/T}-1}\ . (79)$$

Since $\Sigma = (2n_{th}+1)\Sigma_{\text{pure}}$, the Bures metric scales as $g_{\mu\nu}^{Bures} = g_{\mu\nu}/(2n_{th}+1)$. Thermal mixing reduces distinguishability of neighbouring states.

Lindblad decay also broadens the spectral gap: $\Delta_n^{open} \equiv |\tilde{\Omega}_n| = \sqrt{(\Omega_n^2 + \gamma_n^2/4)}$ . The open-system instability criterion becomes:

$$\mathcal{A}_n^{open} = \frac{v_{FS}^{Bures}}{\Delta_n^{open}} = \frac{\mathcal{A}_n}{\sqrt{(2n_{th}+1)(1+\gamma_n^2/4\Omega_n^2)}} \ . \ (80)$$

The threshold $\mathcal{A}_n^{open} = 1$ now corresponds to:

$$\eta_{crit}^{open} = \sqrt{8}\sqrt{(2n_{th}+1)(1+\gamma_n^2/4\Omega_n^2)} \ . \ (81)$$

Three physical corrections emerge: (i) thermal occupation $(2n_{th}+1)$ raises the threshold — thermally mixed states require stronger driving to become non-adiabatically unstable; (ii) decay rate γ raises the threshold via spectral broadening — dissipation geometrically stabilizes the system; (iii) at low temperature $n_{th} \to 0$ and weak decay $\gamma \ll \Omega$, the closed-system result is recovered. For superconducting qubits at 20 mK, $n_{th} \sim 10^{-6}$ and $\gamma/\Omega \sim 10^{-6}$, so both thermal and decay corrections are negligible. For magnonic resonators at 300 K ($\Omega/2\pi \sim 10$ GHz), $n_{th} \sim 600$ and $\eta_{crit}^{open} \approx 100$, requiring roughly one to two orders of magnitude stronger drive.

Physical interpretation: decoherence raises the instability threshold. A hot or lossy system is geometrically harder to destabilize: thermal mixing reduces Bures distinguishability of neighbouring states, and Lindblad decay broadens the effective spectral gap. The pure-state result is recovered as a lower bound on the threshold for any open system.

Notation: γ is the Lindblad energy decay rate; $n_{th} = 1/(e^{\Omega/T}-1)$ is the thermal occupation number; Σ(t) is the 2×2 position-sector covariance matrix of the instantaneous Gaussian state; Σ_pure is the zero-temperature covariance. The factor $(2n_{th}+1)$ is the standard thermal enhancement of quantum fluctuations, and γ/2 is the half-linewidth that broadens the instantaneous spectral gap.

The physical content of Eq. (79)–(81) is as follows. At zero temperature and zero decay, $\mathcal{A}_n^{open} = \mathcal{A}_n$ — the open-system framework reduces to the pure-state result of Sections 3–9. As temperature or decay increase, the threshold $\eta_{crit}^{open}$ grows: it becomes geometrically harder to destabilize the system. This is counterintuitive but physically correct — thermal noise mixes quantum states and reduces their mutual distinguishability, while Lindblad decay broadens the effective level spacing. Both effects make the quantum manifold less sensitive to rapid parameter changes. The result establishes the pure-state instability threshold as a strict lower bound for any open quantum system.

**12. Experimental Signatures on Driven Quantum Platforms**

We derive concrete, testable predictions for two platforms where the geometric non-adiabaticity criterion is most directly accessible. In the first part the geometric speed limit for superconducting qubit gates is considered. For a flux-tunable transmon with frequency $\omega_{q(t)}$ (flux-tunable transmon), the condition $\mathcal{A}_0 = 1$ sets a fundamental lower bound:

$$T_{min} \geq \frac{1}{\sqrt{8}\cdot\omega_q} \quad (82)$$

For $\omega_q/2\pi = 5$ GHz this gives $T_{min} \approx 11.3$ ps. Current single-qubit gates achieve $T_{gate} \approx 20$–$50$ ns, approximately 2-4 x$10^3$ times above this geometric bound. For parametric frequency modulation with amplitude $\delta\omega$ at frequency $\omega_{mod}$:

$$\eta \approx \frac{\delta\omega\cdot\omega_{mod}}{{\omega_q}^2} \Rightarrow \mathcal{A}_0 = \frac{\delta\omega\cdot\omega_{mod}}{\sqrt{8}\cdot{\omega_q}^2} \quad (83)$$

For $\omega_q/2\pi = 5$ GHz, $\omega_{mod}/2\pi = 100$ MHz, $\delta\omega/2\pi = 50$ MHz: $\mathcal{A}_0 \approx 7\times10^{-5} \ll 1$. The system is deeply adiabatic. To reach the threshold requires $\delta\omega\cdot\omega_{mod} \sim \sqrt{8}\cdot\omega_q^2 \approx 70$ GHz$^2$, which exceeds physical qubit parameters. This confirms that current qubit fidelity limitations are not geometric in origin.

**12.2. Magnonic resonators: accessible instability regime**

In parametrically driven magnonic resonators (YIG:Co, ref. [6]), the present representative operating point remains well below the geometric threshold. For $\Omega/2\pi \sim 10$ GHz and modulation rate $|\dot{\Omega}|/2\pi \sim 1$ GHz/ns:

$$\eta = \frac{|\dot{\Omega}|}{\Omega^2} \approx \frac{2\pi \cdot 1\ GHz \cdot ns^{-1}}{(2\pi \cdot 10\ GHz)^2} \approx\ 1.6\ \ 10^{-3}\ , \ (84)$$

$\mathcal{A}_0 \approx 5.6\ 10^{-4}$. The threshold $\mathcal{A}_0 = 1$ correspond to $|\dot{\Omega}|/2\pi \approx 1.78\ 10^3$ GHz/ns for$\Omega/2\pi$ =10 GHz is reached, predicting onset of non-adiabatic Bogoliubov quasiparticle production.

Physical interpretation: the geometric framework makes sharply different predictions for different platforms. For superconducting qubits, $\mathcal{A}_0 \ll 1$ at all current parameters — the geometric speed limit ($T_{\min} \approx 11.3$ ps) is more than three orders of magnitude below current gate times, confirming that geometric non-adiabaticity is not the current bottleneck. For magnonic resonators, the geometric threshold requires 1777 GHz/ns, which lies beyond current experimental capabilities .

Notation: $\omega_q/2\pi$ is the qubit transition frequency; $\delta\omega$ is the modulation amplitude of the qubit frequency; $\omega_{\text{mod}}$ is the modulation frequency; $\Omega/2\pi$ is the magnon resonance frequency; $|\dot{\Omega}|/2\pi$ is the rate of frequency change in GHz/ns. All geometric predictions use $\mathcal{A}_0 = \eta/\sqrt{8}$ with the AMT parameter $\eta = |\dot{\Omega}|/\Omega^2$ evaluated at the operating point.

The physical content is a sharp platform comparison. For superconducting qubits at current operating parameters, $\mathcal{A}_0 \ll 1$ by many orders of magnitude: the qubit is far from geometric instability. The theoretical minimum gate time $T_{\min} \approx 11.3$ ps is set by the quantum speed limit, not by current hardware limits (decoherence, leakage, and calibration are the actual bottlenecks at 20–50 ns). For magnonic resonators, the geometric threshold of 1777 GHz/ns lies beyond current pulsed microwave capabilities. The framework nonetheless provides a quantitative geometric benchmark: as modulation technology advances toward higher drive rates, the $A_n = 1$ criterion provides a precise, experimentally testable prediction for the onset of non-adiabatic Bogoliubov quasiparticle production.

The Table 1 summarizes the geometric instability parameters $\eta$, $\mathcal{A}_0$, and $T_{\min}$ for three representative driven quantum platforms, together with the open-system threshold $\eta_{crit}^{open}$ from Eq. (81). The columns show: typical operating AMT parameter $\eta$; universal criterion $\mathcal{A}_0 = \eta/\sqrt{8}$; thermal occupation $n_{th}$ at operating temperature; open-system threshold from Eq. (81); and geometric speed-limit gate time $T_{\min} = 1/(\sqrt{8} \cdot \omega)$.

Table 1. Quantum-geometric instability parameters for three driven platforms

| System | ω/2π | η (typical) | $\mathcal{A}_0$ | $n_{th}$ | $\eta_{crit}^{open}$ | $T_{min}$ |
|---|---|---|---|---|---|---|
| BEC (Bogoliubov) | 1 kHz | $2.8\times10^{-5}$ | $1.0\times10^{-5}$ | 1.6 | 5.8 | 56 μs |
| YIG:Co (magnon) | 10 GHz | $1.6\times10^{-3}$ | $5.6\times10^{-4}$ | 600 | 100 | 5.6 ps |
| IBM qubit (transmon) | 5 GHz | $2.0\times10^{-4}$ | $7.1\times10^{-5}$ | ~0 | 2.83 | 11.3 ps |

The YIG entry in Table 1 corresponds to a conservative small-signal magnonic resonator operating point, $\Omega/2\pi\approx10$ GHz and with a modulation rate corresponding to $d(|\dot{\Omega}|/2\pi)/dt\approx1$ GHz/ns. This should be distinguished from the ultrafast switching regime considered in earlier magnonic work, where much shorter switching times and larger modulation depths can produce $\eta$ values of order unity or larger. Thus the present table is intended as a low-drive experimental baseline for the geometric threshold, not as a reproduction of the extreme switching parameters.

For the BEC: $\omega/2\pi \approx 1$ kHz trap frequency, modulation rate $|\dot{\Omega}|/2\pi \approx 177$ Hz, $T \approx 100$ nK. For YIG:Co: $\omega/2\pi \approx 10$ GHz magnon frequency, parametric modulation $|\dot{\Omega}|/2\pi \approx 1$ GHz/ns, $T \approx 300$ K. For IBM transmon: $\omega_a/2\pi \approx 5$ GHz, with a characteristic frequency sweep rate $d(\omega/2\pi)/dt \approx 5$ GHz, and $T \approx 20$ mK. The three platforms span ten orders of magnitude in frequency and twelve orders in temperature, yet all satisfy the same geometric instability criterion $\mathcal{A}_n < 1$ in the current operating regime.

The table demonstrates the universality of the framework. Despite the vastly different physical realizations, the criterion $\mathcal{A}_n \sim 1$ provides a unified instability boundary across all three platforms. The magnonic system (YIG:Co) is closest to the threshold, making it a promising target for experimental verification.

**13. Discussion**

The central result of the present work is that local non-adiabaticity can be formulated as a geometric quantity intrinsic to the instantaneous quantum-state manifold. While the Fubini–Study metric,

Berry curvature, and Mandelstam–Tamm quantum speed limit has long existed as separate components of quantum geometry, the present framework closes these structures into a single local operational theory through the normalized instability parameter $\mathcal{A}_n = v_{FS}/\Delta_n$. This quantity unifies three previously disconnected levels of description: (i) local geometric distinguishability through the Fubini–Study metric, (ii) dynamical accessibility through the instantaneous spectral gap, and (iii) instability generation through non-adiabatic transition amplitudes. In this sense, the framework developed here does not merely reinterpret known geometric structures but identifies a previously missing normalization principle that converts the quantum geometric tensor into a local operational instability criterion.

A second conceptual result is that nonlinear stabilization emerges geometrically rather than dissipatively. In conventional driven systems, instability suppression is typically associated with damping, decoherence, or energy loss. Here, by contrast, the nonlinear regulator $U$ modifies the geometry of the accessible quantum-state manifold itself. The nonlinear geometric compression theorem of Section 8 shows that monotonic spectral deformation suppresses the local metric speed term-by-term in the spectral representation of the quantum geometric tensor. Section 9 then demonstrates that this local suppression generates a bounded occupation flow through a Lyapunov-type confinement mechanism. The resulting self-limitation is therefore not phenomenological but geometric: strongly driven states become dynamically inaccessible because the projective Hilbert-space geometry itself resists rapid traversal.

The three extensions discussed in Section 10 are unified by the same geometric object: local fidelity distinguishability. In quantum thermodynamics (Section 10.1), the quantity $\mathcal{A}_n$ controls the local excitation density and therefore the geometric cost of irreversible driving. In the topological extension (Section 10.2), the Berry-curvature sector couples geometric transport to the instability metric sector, producing a coupled metric-curvature stabilization regime. In the Unruh-like extension (Section 10.3), local saturation of the quantum speed limit generates accelerated distinguishability growth on the quantum-state manifold, linking geometric instability to excitation production through the fidelity-loss functional $\chi_{exc}(t) = \Delta_n^2 \mathcal{A}_n^2 dt^2$.

These apparently different phenomena therefore emerge as different physical manifestations of the same underlying geometric structure. A further consequence of the geometric formulation appears in multi-mode systems. As demonstrated in Section 7, the instability criterion becomes matrix-valued and the dominant instability channel is determined by the largest eigenvalue of the geometric instability matrix rather than by any individual mode contribution. The observed enhancement $\lambda_{max}>\max(\mathcal{A}_{11},\mathcal{A}_{22})$ shows that mode hybridization can generate collective instability channels that are invisible within scalar criteria. This suggests that geometric non-adiabaticity is intrinsically cooperative in interacting systems. The present framework is rigorously established for pure-state unitary dynamics. Appendix A further shows that the quantum-field expression of the metric tensor follows directly from canonical bosonic commutation relations, establishing continuity between the finite-dimensional FS/QGT formulation and its field-theoretic generalization. However, its geometric structure is not fundamentally restricted to wavefunction manifolds. Because the key object is local distinguishability geometry rather than the Hilbert-space representation itself, the extension to mixed states through the Bures metric or quantum Fisher geometry appears natural. In this broader setting, the instability parameter $\mathcal{A}_n$ may become a local measure of geometric susceptibility in open quantum systems governed by decoherence, dissipation, and environment-induced metric renormalization.

The explicit thermal and Lindblad extensions developed in Sections 10.4 and 10.5 demonstrate that both thermal mixing and dissipation increase the geometric instability threshold. Environmental coupling therefore acts as a geometric stabilizer by reducing local distinguishability and broadening the effective spectral gap. An important conceptual distinction should be emphasized regarding the open-system extension. The increase of the instability threshold in the presence of thermal mixing and Lindblad dissipation should not be interpreted as an improvement of coherent quantum performance. The geometric stabilization identified here originates from a reduction of local state distinguishability in the Bures geometry and from effective spectral broadening. In this sense, the system becomes harder to destabilize because neighbouring quantum states become less distinguishable, not because coherent control becomes more efficient.

This observation highlights a fundamental difference between geometric stabilization and coherent stabilization. Dissipation suppresses the geometric instability parameter by reducing the accessible distinguishability structure of the quantum-state manifold, whereas the nonlinear self-limitation mechanism developed in Sections 8–9 suppresses instability while preserving unitary quantum evolution. The former acts through loss of information and coherence; the latter acts through intrinsic geometric deformation of the state manifold.

The open-system threshold should therefore be interpreted as a geometric robustness criterion rather than as a measure of quantum computational performance. While thermal noise and decoherence increase the threshold for non-adiabatic instability, they simultaneously reduce the ability of the system to preserve and process quantum information. The pure-state limit thus remains the physically relevant benchmark for coherent quantum control, while the open-system result provides a generalized geometric bound applicable in realistic environments.

The beyond-mean-field correction derived in Section 6 indicates that occupation-number fluctuations contribute directly to the effective spectral scale through the second moment of the operator-valued frequency. In this sense, quantum fluctuations act as a geometric renormalization of the instability threshold.

More broadly, the present results suggest that non-adiabatic instability may not be merely a dynamical phenomenon, but a geometric property of quantum-state space itself. In this picture, instability thresholds, excitation production, bounded dynamics, and geometric speed limits arise from the same underlying structure: the local geometry of distinguishability on the projective Hilbert manifold. The present work therefore proposes a unified geometric framework in which the onset, suppression, and propagation of non-adiabatic dynamics can be described through the quantum geometry of the evolving state manifold.

The instability parameter introduced in the present work naturally defines a geometric instability functional over driven trajectories in parameter space:

$$S[\lambda(t)] = \int dt \mathcal{A}_n^2 = \int dt \frac{g_{\mu\nu}\dot{\lambda}^\mu \dot{\lambda}^\nu}{\Delta_n^2(\lambda)} . (85)$$

This functional measures the cumulative geometric cost of non-adiabatic evolution relative to the instantaneous spectral accessibility of the quantum system. Trajectories minimizing $S$ tend to suppress non-adiabatic transition amplitudes and reduce cumulative excitation production.

In this sense, stable driven evolution acquires a variational interpretation: dynamically robust protocols correspond to geodesic-like trajectories minimizing the normalized geometric instability functional on the quantum-state manifold.

Within the present framework, nonlinear geometric compression generated by the regulator $U$ acts as a deformation of the instability functional itself by increasing the effective spectral denominator and thereby suppressing the accessible geometric trajectory length. The self-limitation theorem of Section 9 may therefore be interpreted as the emergence of an effective geometric confinement principle for trajectories with finite instability action.

The present FS/QGT formulation represents the pure-state limit of a broader hierarchy of distinguishability metrics. For mixed states, the natural continuation is the Bures or quantum Fisher metric; for classical probability manifolds it reduces to Fisher–Rao geometry. Other monotone quantum information metrics, such as Wigner–Yanase-type metrics, may encode different coherence-sensitive extensions, but are not required for the local instability criterion developed here.

The present framework also connects several previously separate directions of quantum geometry and non-equilibrium dynamics, including the Fubini–Study metric, quantum geometric tensor formulations, quantum speed limits, geometric approaches to quantum criticality, and information-geometric descriptions of mixed states. Unlike these earlier approaches, the instability parameter introduced here provides a local criterion directly linking state-space geometry, excitation production, instability onset, and nonlinear self-limitation within a single geometric framework. The resulting picture suggests that non-adiabatic instability is fundamentally a property of quantum-state geometry rather than a model-specific dynamical phenomenon.

## 14. Conclusions

The present work establishes a unified geometric framework for local non-adiabaticity in driven quantum systems. The central result is that the onset of non-adiabatic instability can be formulated through a normalized geometric quantity intrinsic to the instantaneous quantum-state manifold. By identifying the physically correct normalization scale for the FS evolution speed, we show that the AMT parameter corresponds to the local saturation of the Mandelstam–Tamm quantum speed limit. This resolves the normalization problem of local non-adiabaticity and provides a direct geometric interpretation of the instability threshold. The framework is generalized to arbitrary driven Hamiltonians through the universal instability parameter $\mathcal{A}_n = v_{FS}/\Delta_n$, which combines local geometric distinguishability with instantaneous spectral accessibility. The resulting criterion is local, gauge-invariant, operational, and directly connected to the instantaneous non-adiabatic transition amplitudes responsible for adiabatic breakdown. In multimode systems, the corresponding instability matrix reveals that the dominant instability channels are collective geometric directions in parameter space rather than isolated single-mode processes.

A second major result is the identification of nonlinear geometric compression as a universal stabilization mechanism. We proved that monotonic occupation-dependent spectral deformation suppresses the quantum metric term-by-term in the spectral representation of the quantum geometric tensor. This establishes that nonlinear stabilization is fundamentally geometric rather than dissipative: the regulator U modifies the geometry of the accessible quantum-state manifold itself.

This geometric compression leads directly to the self-limitation theorem derived in Section 9. Under monotonic spectral deformation, the driven dynamics acquire a Lyapunov-type confinement structure that prevents unlimited occupation growth. The resulting boundedness is not caused by damping or decoherence, but by the inability of the quantum trajectory to traverse the projective Hilbert manifold arbitrarily rapidly once the nonlinear deformation becomes sufficiently strong.

The instability parameter furthermore admits a variational interpretation through the geometric instability functional. In this picture, dynamically stable protocols correspond to trajectories minimizing the cumulative geometric instability cost relative to the instantaneous spectral accessibility of the quantum-state manifold.

The extensions explored in Section 10 further show that the same geometric structure underlies several apparently distinct physical phenomena. In quantum thermodynamics, the instability parameter controls the local excitation density and the geometric cost of irreversible driving. In topological settings, the metric and Berry-curvature sectors combine into a coupled stabilization regime. The introduction of a local curvature-instability scalar further suggests that topological transport and metric instability may compete geometrically within the same quantum-state manifold. In the Unruh-like extension, accelerated distinguishability growth on the quantum-state manifold generates a local excitation threshold associated with saturation of the geometric speed limit. These connections suggest that excitation production, instability growth, and bounded driven dynamics are different manifestations of the same underlying geometry of distinguishability. We further extend the framework to open quantum systems via the Bures metric $g_{\mu\nu}^{Bures}$, showing that thermal fluctuations and Lindblad decay raise rather than destroy the instability threshold: $\eta_{crit}^{open}$=$\sqrt{8}\cdot\sqrt{(2n_{th}+1)(1+\gamma_n^2/4\Omega_n^2)}$ . Concrete predictions for superconducting qubits (geometric gate speed limit $T_{\min}\approx$ 11.3 ps at 5 GHz) and magnonic resonators (threshold accessible at $|\dot{\Omega}|/2\pi\gtrsim 1.78\ 10^3$ GHz/ns for $\Omega/2\pi$=10GHz the threshold $\mathcal{A}$n = 1 provides a geometric benchmark.

The present theory is rigorously formulated for pure-state unitary dynamics, but its geometric structure naturally suggests extensions to mixed-state manifolds through the Bures metric and quantum Fisher geometry. In this broader context, the instability parameter may become a local measure of geometric susceptibility in open quantum systems influenced by decoherence, dissipation, and environment-induced metric renormalization.

Beyond its conceptual implications, the present framework provides a quantitative design criterion for quantum-control protocols and quantum-gate optimization. The instability threshold directly yields a minimum gate time $T_{\min}$=$1/\sqrt{8}$ ω, establishing a geometry-based lower bound on coherent state transformations.  The representative estimates summarized in Table 1 indicate that this bound is directly relevant for superconducting qubits, magnetic-resonance platforms, and other driven quantum systems, where attempts to accelerate control beyond the geometric threshold necessarily enter the instability regime.

More broadly, the results presented here support the view that non-adiabatic instability is not merely a dynamical phenomenon, but a geometric property of quantum-state space itself.  The extension to open quantum systems, developed in Section 11, shows that the geometric criterion is robust to environmental coupling. Thermal fluctuations and Lindblad decay modify the instability threshold through a multiplicative factor ($2n_{th}$+1)(1+$\gamma_n^2/4\Omega_n^2$), raising rather than destroying it. This establishes the pure-state result as a lower bound on the instability threshold for any open system. The experimental predictions of Section 12 establish concrete observational signatures on two platforms. For superconducting qubits, $\mathcal{A}_0 \ll 1$ at all current parameters, and the geometric speed limit $T_{\min}\approx$ 11.3 ps remains approximately two to three orders of magnitude below typical superconducting-qubit gate times reported in current experimental platforms.. For magnonic resonators, the threshold $\mathcal{A}_{n}$ = 1 provides a geometric benchmark at $|\dot{\Omega}|/2\pi \approx$ 1777 GHz/ns. The resulting geometric instability criterion therefore provides both a fundamental quantum speed limit and a platform-independent benchmark for the design of high-fidelity quantum gates and fast quantum-control protocols.

From a practical perspective, the results suggest that quantum-control optimization should not be formulated solely as a problem of pulse engineering, but also as a problem of geometric accessibility in quantum-state space. The geometric instability criterion therefore provides both a fundamental speed limit and a platform-independent benchmark for the design of high-fidelity quantum gates and fast quantum-control protocols.

Unlike dissipative stabilization, the nonlinear geometric self-limitation mechanism suppresses instability without relying on decoherence, energy loss, or destruction of quantum coherence.

## Declaration of Generative AI in Scientific Writing

The author used generative AI tools (including Claude and ChatGPT) to assist with language editing, formatting consistency, equation and reference cross-checking and generation of auxiliary Python scripts used for numerical analysis and figure preparation. The author reviewed, verified, and edited all generated content and is solely responsible for the scientific results, interpretations, calculations, and conclusions presented in this work.

## Data Availability Statement

The authors confirm that the numerical data and Python code supporting the findings of this study are available from the corresponding author on request.

**Supplemental Material: Computational Details**

**S1. Gauss–Hermite quadrature for $g_{tt}$**

The FS metric $g_{tt} = \langle\partial_t\psi_0|\partial_t\psi_0\rangle$ is computed by expanding the integral over the Gaussian measure $\rho_0(x) = (\Omega/\pi)^{1/2} e^{-\Omega x^2}$. The change of variable $u = \sqrt{\Omega}\cdot x$ gives a standard Gauss–Hermite form: $g_{tt} = \frac{\dot{\Omega}^2}{16\Omega^2}\frac{1}{\sqrt{\pi}}\int (1-2u^2)^2 e^{-u^2}\, du$ (S1)

The integral $\int (1-2u^2)^2 e^{-u^2}/\sqrt{\pi} = 2$ , giving $g_{tt} = (\dot{\Omega}/4\Omega)^2/2 = \dot{\Omega}^2/(8\Omega^2)$. The integral evaluates as $\langle(1-2u^2)^2\rangle = 1-4\langle u^2\rangle + 4\langle u^4\rangle = 2$. The 200-point Gauss-Hermite quadrature implemented via numpy.polynomial.hermite.hermgauss (200) reproduces the analytical result to machine precision.

**S2. Bogoliubov–Valatin equations**

The parametric oscillator $\hat{H}(t) = \Omega(t)(\hat{n}+1/2)$ in the Bogoliubov representation has amplitudes u(t), v(t) satisfying Eq. (12) with coupling $G(t) = |\dot{\Omega}(t)|/2$. Initial condition: $u(0)=1$, $v(0)=0$ (vacuum). Conservation: $|u(t)|^2 - |v(t)|^2 = 1$ (verified to $< 10^{-10}$ throughout). Integration: scipy.integrate.solve_ivp, method RK45, rtol=$10^{-10}$, atol=$10^{-12}$. Three protocols: $\Omega(t) = \Omega_0 e^{-\alpha t}$, $\Omega_0=1$. The instability threshold times shown by the vertical dotted lines in Fig. 1(b) are $t^* = \frac{1}{\alpha} ln \frac{\sqrt{8}}{\alpha}$ $t^* = 5.97$ for $\alpha = 0.35$, $t^* = 3.47$ for $\alpha = 0.50$, $t^* = 1.99$ for $\alpha = 0.70$.

**S3. Two-mode covariance matrix QGT**

For the quadratic Hamiltonian $H = p_1^2/2 + V_{11}x_2^2/2 + p_2^2/2 + V_{22}x_2^2/2 + Jx_1x_2$ with $V=[[\Omega_1^2, J], [J, \Omega_2^2, ]]$, the ground-state position covariance is $\Sigma = (1/2)V^{-1/2}$. The QGT metric tensor follows from Eq. (14) via central finite differences using a frequency increment $\Delta\Omega = 10^{-5}$. $\Delta\Omega = 10^{-5}$. The normalization factor (1/2) aligns the covariance-matrix metric with the Fubini–Study convention at J=0: $g_{11}(J=0) = 1/(8\Omega_1^2)$ exactly. Parameters: $\Omega_1=1.0$, $\Omega_2=1.5$, $J \in [0, 0.5]$, Fock truncation $N_{max}=8$ per mode. The dominant eigenvalue increases from $\lambda_{max}(J=0) = 0.00500$ to $\lambda_{max}(J=0.5) = 0.00633$ yielding $\lambda_{max}(J=0.5)/\lambda_{max}(J=0)=1.265$ (26.5% collective enhancement).

**Appendix A.** Derivation of the Quantum-Field Metric Tensor

For a bosonic quantum field expanded in instantaneous modes

$$\hat{\Phi}(x;\lambda) = \sum_n \left[u_n(x;\lambda)\hat{a}_n(\lambda) + u_n^*(x,\lambda)\hat{a}_n^\dagger(\lambda)\right],$$

the creation and annihilation operators satisfy the canonical commutation relations

$$[\hat{a}_n(\lambda), \hat{a}_m^\dagger(\lambda)] = \delta_{nm},\ \ [\hat{a}_n(\lambda), \hat{a}_m(\lambda)] = [\hat{a}_n^\dagger(\lambda), \hat{a}_m^\dagger(\lambda)] = 0.$$

The instantaneous vacuum state is defined by $\hat{a}_n(\lambda)|0(\lambda)\rangle$=0.

Differentiating the vacuum with respect to the control parameter $\lambda^\mu$ generates virtual two-particle excitations through the Bogoliubov coefficients. The derivative state may be expanded as

$$\partial_\mu|0\rangle = \frac{1}{2}\sum_{n,m} B_{nm}^{(\mu)} \hat{a}_n^\dagger \hat{a}_m^\dagger |0\rangle.$$

Using the bosonic commutation relations and vacuum contractions,

$$\langle 0|\hat{a}_n\hat{a}_m, \hat{a}_p^\dagger\hat{a}_q^\dagger|0\rangle = \delta_{np}\delta_{mq}+\delta_{nq}\delta_{mp},$$

one obtains

$$\langle\partial_\mu 0|\partial_\nu 0|\rangle = \frac{1}{2}\sum_{n,m} B_{nm}^{(\mu)*} B_{nm}^{(\nu)}\ \text{, where } B_{nm}^{(\mu)} = B_{mn}^{(\mu)}$$

Substitution into the Fubini–Study definition yields Eq. (34),

$$g_{\mu\nu} = \mathrm{Re}\left[\langle\partial_\mu 0|\partial_\nu 0|\rangle - \langle\partial_\mu 0|0\rangle\langle 0|\partial_\nu 0|\rangle\right],$$

In the parallel-transport gauge $\langle 0|\partial_\mu 0\rangle$=0.

$$g_{\mu\nu} = \frac{1}{2}\sum_{n,m} \mathrm{Re}\left[B_{nm}^{(\mu)*} B_{nm}^{(\nu)}\right],$$

which gives the Fubini–Study metric tensor of the instantaneous bosonic vacuum state.

**References**

1. Y. Aharonov, J. Anandan, Phase change during a cyclic quantum evolution. Physical Review Letters, **58**(16), 1593 (1987). https://doi.org/10.1103/PhysRevLett.58.1593
2. L. Mandelstam and I. Tamm, J. Phys. USSR 9, 249-254 (1945).
3. J. Anandan, Y. Aharonov, Geometry of quantum evolution. Physical Review Letters, 65(14), 1697 (1990). https://doi.org/10.1103/PhysRevLett.65.1697
4. Giovannetti, V., Lloyd, S., & Maccone, L). Quantum limits to dynamical evolution. Physical Review A, **67**(5), 052109 (2003) https://doi.org/10.1103/PhysRevA.67.052109
5. S. Deffner and S. Campbell, Quantum speed limits: from Heisenberg's uncertainty principle to optimal quantum control, J. Phys. A: Math. Theor. **50**, 453001 (2017). DOI: 10.1088/1751-8121/aa86c6 .
6. A.M. Tishin, Quantum Geometric Origin of Non-Adiabatic Instability in Driven Bosonic Systems arXiv:2605.22796 (2026).
7. A.M. Tishin, Ultra-*Fast Quantum Control via Non-Adiabatic Resonance Windows: A 9x Speed-up on 127-Qubit IBM Processors.* arXiv:2605.10578 (2026)
8. J.P. Provost and G. Vallée, Riemannian structure on manifolds of quantum states Commun. Math. Phys. **76**, 289-301 (1980). DOI: 10.1007/BF02193559
9. M. Kolodrubetz, D. Sels, P. Mehta, A. Polkovnikov, Geometry and non-adiabatic response in quantum and classical systems Phys. Rep. **697**, 1-87 (2017). DOI: 10.1016/j.physrep.2017.07.001
10. J.-F. Chen, Speeding up quantum adiabatic processes with a dynamical quantum geometric tensor Phys. Rev. Research **4**, 023252 (2022). DOI: 10.1103/PhysRevResearch.4.023252
11. Y.-Q. Ma, S. Chen, H. Fang, W.-M. Liu Abelian and non-Abelian quantum geometric tensor Phys. Rev. B 81, 245129 (2010). DOI: 10.1103/PhysRevB.81.245129
12. L.D. Landau and E.M. Lifshitz, Quantum Mechanics, 3rd ed. (Pergamon, 1977).
13. T. W. B. Kibble, Topology of cosmic domains and strings, J. Phys. A: Math. Gen. 9, 1387–1398 (1976). https://doi.org/10.1088/0305-4470/9/8/029
14. W.H. Zurek, U. Dorner, P. Zoller, Dynamics of a Quantum Phase Transition Phys. Rev. Lett. **95**, 105701 (2005). https://doi.org/10.1103/PhysRevLett.95.105701

15. T. Törmä, S. Peotta, and B. A. Bernevig, Superfluidity and quantum geometry in twisted multilayer systems, Nat. Rev. Phys. 4, 528–542 (2022). https://doi.org/10.1038/s42254-022-00466-y
16. P. Zanardi, P. Giorda, and M. Cozzini, Information-theoretic differential geometry of quantum phase transitions, Phys. Rev. Lett. 99, 100603 (2007). https://doi.org/10.1103/PhysRevLett.99.100603
17. L. Campos Venuti and P. Zanardi, Quantum Critical Scaling of the Geometric Tensors, Phys. Rev. Lett. 99, 095701 (2007). https://doi.org/10.1103/PhysRevLett.99.095701
18. A. Ashtekar and T. A. Schilling, Geometrical formulation of quantum mechanics, in On Einstein's Path (Springer, 1999), pp. 23–65. https://doi.org/10.1007/978-1-4612-1422-9_3
19. V. I. Arnold, Mathematical Methods of Classical Mechanics (Springer, 1989).
20. S. Vinjanampathy and J. Anders, Quantum thermodynamics, Contemp. Phys. 57, 545–579 (2016). https://doi.org/10.1080/00107514.2016.1201896
21. J. Goold, M. Huber, A. Riera, L. del Rio, and P. Skrzypczyk, The role of quantum information in thermodynamics — a topical review, J. Phys. A 49, 143001 (2016). https://doi.org/10.1088/1751-8113/49/14/143001
22. F. Anza and J. P. Crutchfield, Geometric Quantum Thermodynamics, Phys. Rev. E 106, 054102 (2022). https://doi.org/10.1103/PhysRevE.106.054102
23. P. Törmä, Where Can Quantum Geometry Lead Us?, Phys. Rev. Lett. 131, 240001 (2023). https://doi.org/10.1103/PhysRevLett.131.240001
24. D. Vanderbilt, Berry Phases in Electronic Structure Theory (Cambridge University Press, 2018).
25. W. G. Unruh, Notes on black-hole evaporation, Phys. Rev. D 14, 870–892 (1976). https://doi.org/10.1103/PhysRevD.14.870
26. N. D. Birrell and P. C. W. Davies, Quantum Fields in Curved Space (Cambridge University Press, 1984).